\documentclass[useAMS,usenatbib]{mn2e}

\usepackage{amsmath}
\usepackage{amssymb}
\usepackage{bm}
\usepackage{color}
\usepackage{mathrsfs}
\usepackage{graphicx}

\usepackage{epstopdf}

\renewcommand{\vec}[1]{\ensuremath{\mathbf{#1}}}

\newcommand{\Myr}{\ensuremath{\,\mathrm{Myr}}}
\newcommand{\Gyr}{\ensuremath{\,\mathrm{Gyr}}}
\newcommand{\kpc}{\ensuremath{\,\mathrm{kpc}}}
\newcommand{\pc}{\ensuremath{\,\mathrm{pc}}}
\newcommand{\kms}{\ensuremath{\,\mathrm{km\ s}^{-1}}}

\newcommand{\kmskpc}{\ensuremath{\,\mathrm{{km\ s}^{-1}\ {kpc}^{-1}}}}

\newcommand{\eq}[1]{\begin{align}#1\end{align}}

\newcommand\aj{{AJ}}%
\newcommand\apj{{ApJ}}%
\newcommand\apjl{{ApJ}}%
\newcommand\apjs{{ApJS}}%
\newcommand\aap{{A\&A}}%
\newcommand\aapr{{A\&A~Rev.}}%
\newcommand\mnras{{MNRAS}}%
\newcommand\nar{{New A Rev.}}%
\newcommand\pasj{{PASJ}}%
\title[Shepherding the Ophiuchus Stream]{Shepherding Tidal Debris with the Galactic Bar: The Ophiuchus Stream}

\author[Hattori, Erkal \& Sanders]{Kohei Hattori$^{1}$\thanks{khattori@ast.cam.ac.uk},
Denis Erkal$^{1}$\thanks{derkal@ast.cam.ac.uk},
Jason L. Sanders$^{1}$\thanks{jls@ast.cam.ac.uk}\\
$^{1}$Institute of Astronomy, University of Cambridge, Madingley Road, Cambridge CB3 0HA, UK}

\begin{document}


\pagerange{\pageref{firstpage}--\pageref{lastpage}} \pubyear{20xx}

\maketitle
\label{firstpage}

\begin{abstract}
The dynamics of stellar streams in rotating barred potentials is explained for the first time. Naturally, neighbouring stream stars reach pericentre at slightly different times. In the presence of a rotating bar, these neighbouring stream stars experience different bar orientations during pericentric passage and hence each star receives a different torque from the bar. These differing torques reshape the angular momentum and energy distribution of stars in the stream, which in turn changes the growth rate of the stream. For a progenitor orbiting in the same sense as the bar's rotation and satisfying a resonance condition,
the resultant stream can be substantially shorter or longer than expected, depending on whether the pericentric passages of the progenitor occur along the bar's minor or major axis respectively. We present a full discussion of this phenomenon focusing mainly on streams confined to the Galactic plane. In stark contrast with the evolution in static potentials, which give rise to streams that grow steadily in time, rotating barred potentials can produce dynamically old, short streams. This challenges the traditional viewpoint that the inner halo consists of well phase-mixed material whilst the tidally-disrupted structures in the outer halo are more spatially coherent. We argue that this mechanism plays an important role in explaining
the mysteriously short Ophiuchus stream that was recently discovered near the bulge region of the Milky Way.

\end{abstract}


\section{Introduction} \label{section:intro}

Recent large-scale deep photometric surveys have revealed that
the distribution of stars in the Galactic halo is not smooth,
but that the stellar halo contains many filamentary substructures dubbed stellar streams (see \citealt{Belokurov2013} and references therein). Since these filaments are thought to be remnants of tidally disrupted systems such as
dwarf galaxies or globular clusters, these stellar streams are also called tidal streams.

The discovery of stellar streams is of importance in Galactic astronomy in two respects.
First, their discovery is evidence that the Milky Way is still growing in mass by accreting smaller systems,
which favours the hierarchical formation scenario of the Milky Way \citep{SZ1978}.
Second,
since a stellar stream is shaped by the tidal forces from the Milky Way,
kinematic information on stellar streams provides constraints on the shape of the Galactic potential
\citep{Johnston1999, Murali1999, Binney2008, Koposov2010, Penarrubia2012, Sanders2014, Bovy2014,Gibbons2014, Bowden2015, Kupper2015, Sanderson2015}. Since the astrometric satellite \emph{Gaia} \citep{Perryman2001}
will provide accurate positions and velocities of a billion stars in the Milky Way in the near future, studies on the dynamics of stellar streams and their use as a probe of the Galactic potential
are particularly timely.

To date, most of the studies on stream dynamics
have been focused on the evolution of stellar streams in static potentials
\citep[e.g.][]{Helmi1999, Tremaine1999, Eyre2011}.
However, the existence of streams itself suggests that the Milky Way mass is increasing in time and hence that
the Galactic potential is not static.
Recently, \cite{Buist2015} studied the evolution of streams in time-dependent spherical potentials
and found that the evolution of the Galactic potential
is imprinted on the morphology of the stellar stream. These results suggest that more general time evolution of the potential should also be important in shaping tidal streams. One obvious contribution to the time-varying potential is provided by the rotating bar towards the centre of the Milky Way so it is natural to ask what impact this rotating quadrupole in the potential has on stream formation.

The effect of the rotating bar is expected to be particularly important
for those stellar streams that pass near the bulge region.
Recently,
\cite{Bernard2014} discovered the Ophiuchus stellar stream in Pan-STARRS1 data near the bulge region of the Milky Way.
Intriguingly, the Ophiuchus stream is only $1.6 \pm 0.3 \kpc$ long with an aspect ratio of $\sim 50$ \citep{Sesar2015}.
Based on orbit fitting in a static axisymmetric potential,
\cite{Sesar2015} estimated that the Ophiuchus stream began disrupting only $\sim 300 \Myr$ ago,
possibly triggered by a disc crossing.
However,
since the ages of the member stars are around $\sim 11.7 \Gyr$ \citep{Sesar2015},
the progenitor system should have experienced $\sim30$ disc-crossings before.
Thus, if we are to adopt their dynamical age,
we need to explain why the progenitor was able to remain unscathed for $\sim 11\Gyr$. On the other hand, if the progenitor of the Ophiuchus stream only entered the potential of the Milky Way recently, it likely entered as part of a larger system whose remnants should still be visible. In this paper we will consider an alternative explanation. We will see that a rotating barred potential can significantly slow the growth rate of a stream that is in resonance with the bar such that long-lived short streams are achievable. This is clearly an important part of the history of the Ophiuchus stream.

This paper is organised in the following manner.
In Section \ref{section:theory},
we provide a qualitative explanation of how a rotating bar affects the evolution of a stream confined to the Galactic plane and show how the stream length is sensitive to the bar pattern speed.
In Section \ref{section:method},
we explain the Lagrange-point stripping method we use to generate mock stellar streams.
In Section \ref{section:2D}, we quantitatively study how streams confined to the equatorial plane grow
in rotating barred potentials using Lagrange-point stripping simulations.
We demonstrate that stellar streams can be shorter or longer than expected
if the progenitor's orbit is in resonance with the bar's rotation.
In Section \ref{section:3D}, we analyse Ophiuchus-like streams
that evolve in rotating barred potentials using Lagrange-point stripping simulations as well as $N$-body simulations.
We compare our $N$-body models with the observed properties of the Ophiuchus stream and find a good match.
In Section \ref{section:discussion},  we discuss how this resonance compares to the well-known Lindblad resonance, as well
as the implications of this mechanism for finding substructures in future surveys. Finally, we conclude in Section \ref{section:conclusion}.

\section{Stream dynamics in rotating barred potentials} \label{section:theory}

We motivated the study of streams in rotating barred potentials by considering the case of the Ophiuchus stream. Before constructing models of the Ophiuchus stream we present a qualitative discussion of the dynamics of streams in rotating barred potentials. We restrict ourselves to orbits which are confined to the plane in which the bar rotates to simplify the discussion. We begin by revising how streams grow in non-rotating potentials.

\subsection{Stream growth in static potentials}

In order to understand how a rotating bar affects a stream, we need to first consider how streams evolve in static potentials. This has been discussed many times previously \citep[e.g.][]{Helmi1999, Eyre2011}. Stream particles arrive at the tidal radius with a distribution of energies and angular momenta. Once released from the progenitor, the effect of the progenitor's gravity is negligible and these particles conserve their energy since the potential is static and conserve the components of their angular momentum corresponding to each rotational symmetry in the potential. At leading order, the orbital frequency of these particles is governed by their energy so particles from each stripping event distribute themselves along the stream, ordered by energy. The energy of the leading tail is lower than the progenitor and the energy of the trailing tail is higher than that of the progenitor. As the stream evolves forwards in time, the distribution of energies in the stream causes it to stretch out with a larger energy spread producing a faster growth.

\subsubsection{Axisymmetric potential (Model A)}
\begin{figure*}
\begin{center}
	\includegraphics[angle=0,width=0.475\columnwidth]{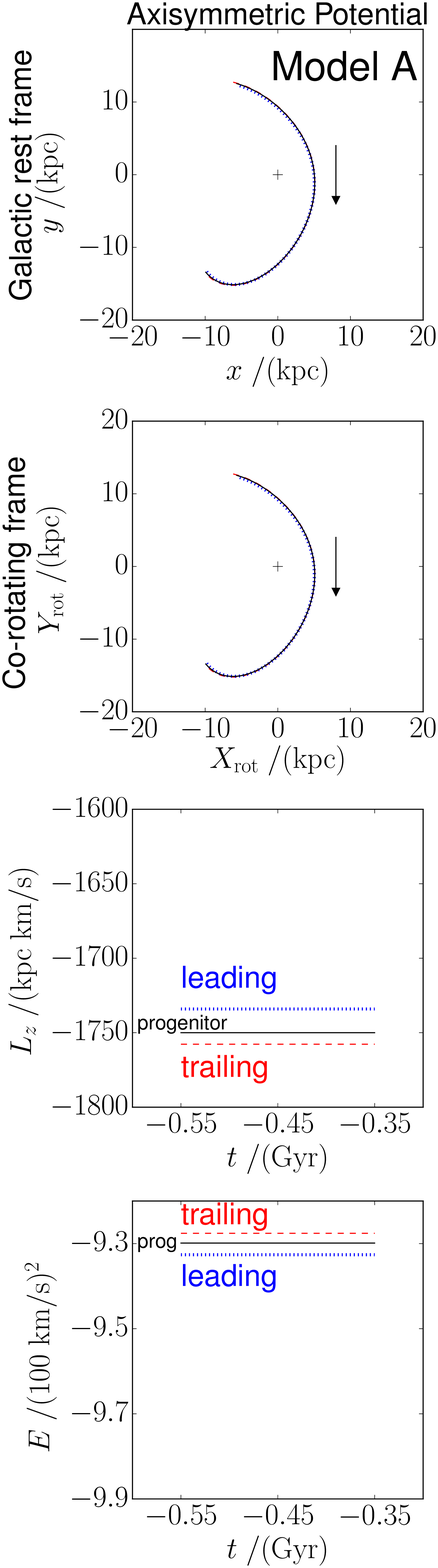}
	\includegraphics[angle=0,width=0.475\columnwidth]{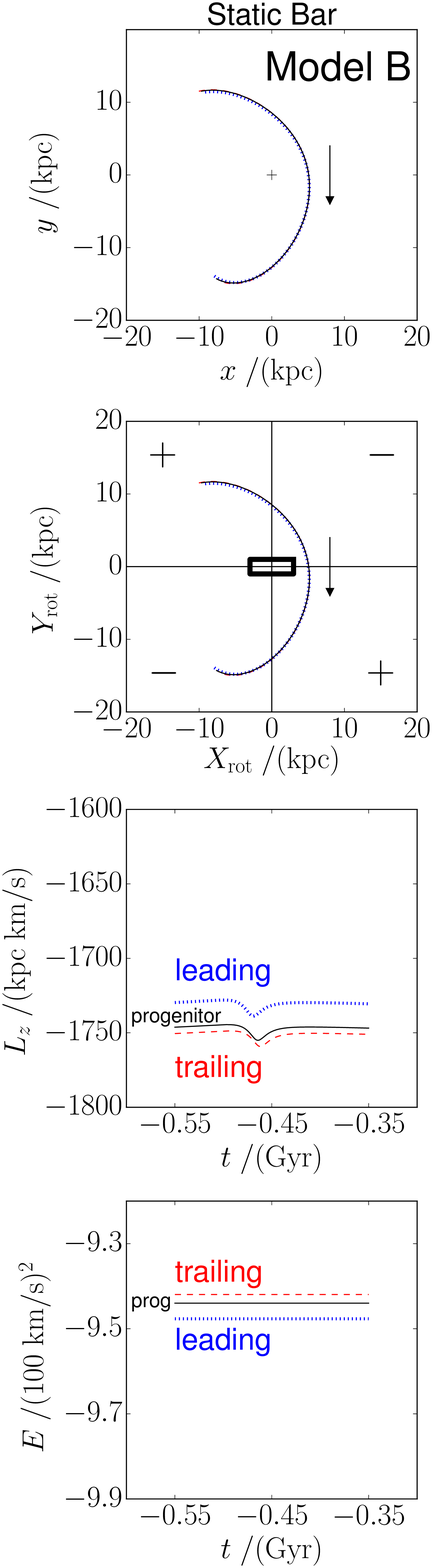}
	\includegraphics[angle=0,width=0.475\columnwidth]{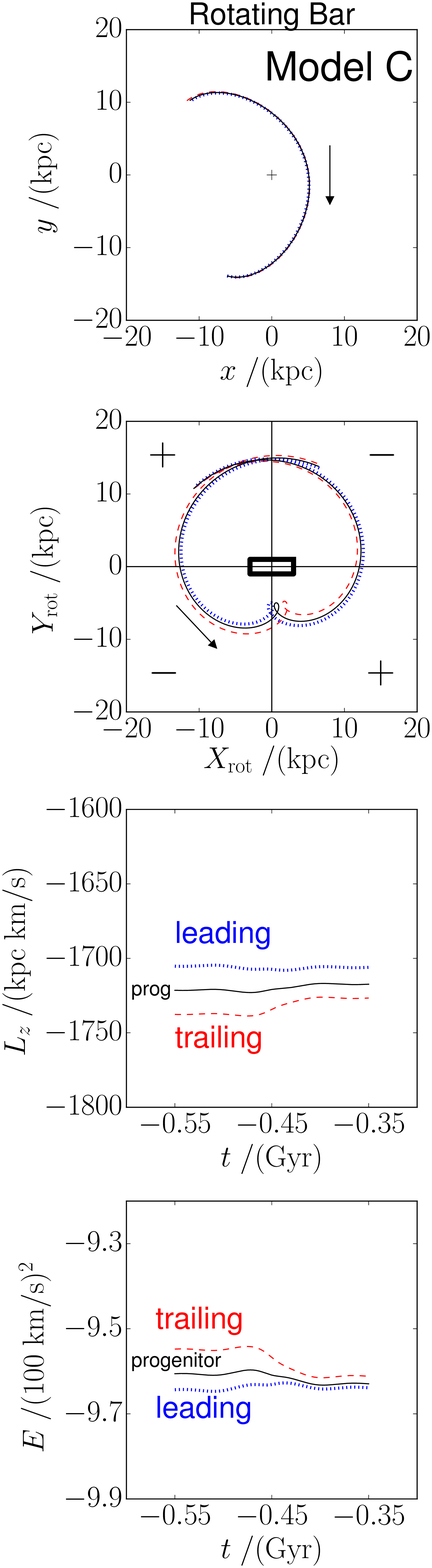}
	\includegraphics[angle=0,width=0.475\columnwidth]{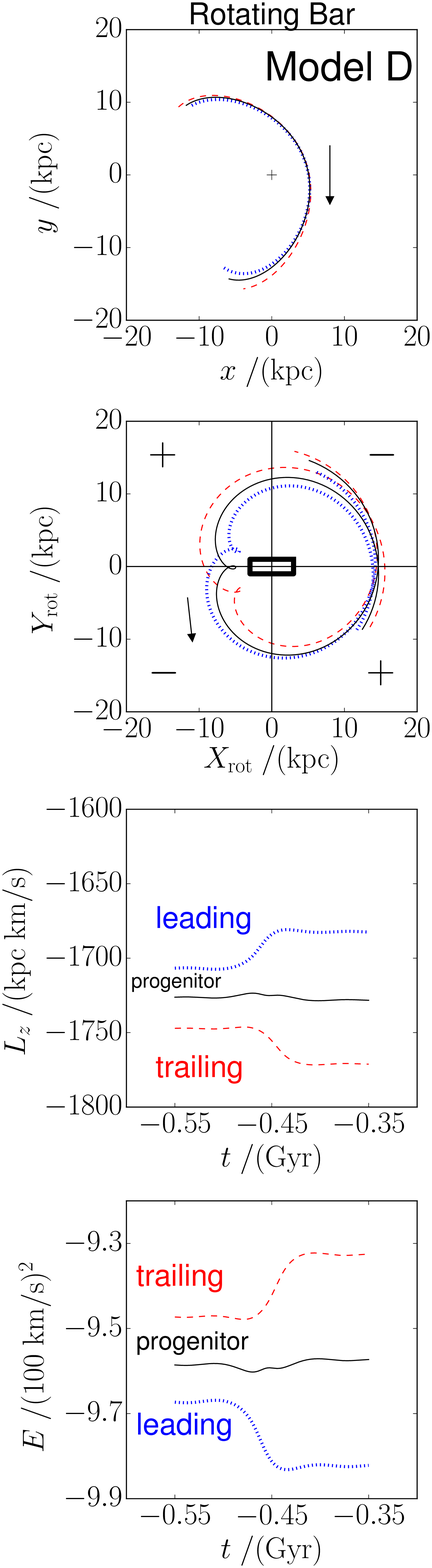}
\end{center}
\caption{
The evolution of the progenitor system (black solid line), a typical star in the leading tail (blue dotted line), and
a typical star in the trailing tail (red dashed line)
near one of the pericentric passages
in the numerical simulations described in Section \ref{section:2D}.
In the four columns,
we show the results for four potential models (Models A, B, C and D),
which are identical to the models in Figure \ref{fig:2D_special_cases}.
In the \textbf{top panels}, we show the orbits in the Galactic rest frame.
In the \textbf{second panels}, we show the orbits in the co-rotating frame such that the horizontal axis is aligned to the bar.
For the barred potentials, the sign of the bar's torque is also shown with `$+$' and `$-$' at each of the quadrants.
In the \textbf{third and fourth panels}, we show the evolution of $L_z$ and $E$, respectively.
}
\label{fig:2D_special_cases_Lz}
\end{figure*}

Now, let us consider a simple model of a stream produced by a progenitor on an eccentric orbit confined to the equatorial plane, $(x,y)$, in an axisymmetric potential. We call this model `Model A'. Our `stream' consists of three particles: a progenitor particle, a leading tail particle and a trailing tail particle. The orbits of these three particles around pericentric passage are shown in the leftmost column of Figure~\ref{fig:2D_special_cases_Lz}. The top panel shows the orbit in the Galactic rest frame and the second panel shows the orbit in a co-rotating frame which is identical since the potential is static. The particles follow roughly the same orbit as their respective energies are very similar. The third and fourth panels show $L_z$ and $E$ which are conserved as expected. The details of these orbits, the stream they are drawn from, and the potential they are evolved in are given in Section \ref{section:2D}. For now, these details are unimportant and Figure \ref{fig:2D_special_cases_Lz} should be thought of as a cartoon.

\subsubsection{Static barred potential (Model B)}

\begin{figure}
\begin{center}
	\includegraphics[angle=0,width=0.95\columnwidth]{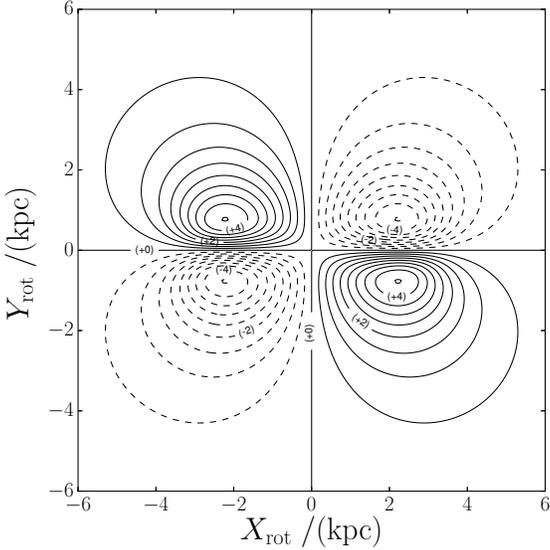}
\end{center}
\caption{
Contours of the $z$-component of the torque in units of $({\rm kpc\;km\;s^{-1} Myr^{-1}})$ from a \protect\cite{Long1992} bar.
The solid-line and dashed-line contours show positive and negative torque, respectively.
The $(X_{\rm rot}, Y_{\rm rot})$-frame is the co-rotating system of the bar
in which the bar is aligned with the $X_{\rm rot}$-axis. The details of the bar are given in Section \protect\ref{section:potential}.
}
\label{fig:torque}
\end{figure}

Next, we consider a potential with a non-rotating bar with its major axis aligned along the $x$ axis. Since the bar is not axisymmetric, it generates a torque in the $z$ direction which changes $L_z$. In Figure \ref{fig:torque} we show contours of the $z$ component of the torque in the disc plane for this bar. The details of the bar model which generates this torque are given in Section \ref{section:2D}. The exact details of this torque are unimportant for the qualitative explanation. We only rely on the sign of the torque in each quadrant, and the fact that the torque decreases as we move away from the bar. Crucially, since the torque falls off with radius, the main change in angular momentum occurs when particles are near pericentre.

Again, we consider a simple stream model confined to the equatorial plane in this potential (Model B). Since this potential is still static, $E$ is conserved. However, $L_z$ is no longer conserved since rotational symmetry is broken. We show the orbit and changes in $L_z$ and $E$ for this case in the second column of panels in Figure \ref{fig:2D_special_cases_Lz}. The second panel (co-rotating frame) also shows the orientation of the bar and the sign of the torque as a reminder that $L_z$ is no longer conserved. As the particles approach pericentre, they receive a torque which first decreases $L_z$ and then increases it, as expected from the signs of the torque. Each particle experiences an almost identical change in angular momentum since they are roughly on the same orbit. Finally, $E$ is conserved since the potential is static. Thus we see that a non-rotating bar affects each particle in the stream almost equally and will not dramatically change the stream growth rate.

\subsection{Stream growth in rotating barred potentials}
Much of the same intuition applies to streams evolving in the presence of rotating bars. When stream particles are far from the bar, their energy is conserved and the stream growth is governed by the distribution of energies along the stream since the potential is effectively static. However, when the stream is near the bar, the potential is strongly time dependent and energy is no longer conserved. This time dependent potential can reshape the energy distribution of the stream and cause its growth rate to increase, decrease, or even become negative.

We consider a simple setup with a potential made up of two parts: a static axisymmetric component and a bar rotating in the $(x,y)$-plane with a pattern speed of $\Omega_{\rm b}$. In this potential, the energy is no longer conserved since the potential is time dependent. Instead, the Jacobi integral,
\eq{ E_{\rm J} = E - \Omega_{\rm b} L_z , \label{eq:jacobi}}
is conserved \citep[see][]{BT2008}. As discussed above, we are interested in how the energy of stream particles is affected by the bar. However, it is conceptually easier to understand how the angular momentum of these particles is affected using the torque. The Jacobi integral allows us to translate the change in angular momentum into a change in energy, and thus into how the bar affects the stream growth rate.

\subsubsection{A short stream in a rotating potential (Model C)}
In the third column of panels in Figure \ref{fig:2D_special_cases_Lz} we consider a simple stream model confined to the $(x,y)$ plane in the presence of a bar rotating at $\Omega_{\rm b}=-58.1\kmskpc$ ( Model C). Now both $E$ and $L_z$ are no longer conserved. In the top panel we see the orbit in the Galactic rest frame looks almost identical to the previous two cases. However, the orbit in the frame co-rotating with the bar now looks dramatically different. Although particles in the stream are roughly on the same orbit, particles in the leading arm arrive first and particles in the trailing arm arrive last. This time difference between particles means the bar rotates and each particle sees the bar in a different orientation. Equivalently, in the co-rotating frame, the orbit of each particle is rotated by the pattern speed times this time delay. Each part of the stream can now experience a different torque. In this example, the progenitor has a pericentre near the minor axis of the bar. The torque on the progenitor is nearly zero during the pericentric passage so $L_z$ is almost conserved. The particle in the leading arm, which arrived at pericentre earlier, receives a predominantly negative torque, decreasing its angular momentum. This decrease in angular momentum translates into an increase in energy via the Jacobi integral. The opposite is true for the particle in the trailing arm which decreases its energy. As a result of this pattern of energy changes, we see that the energy spread between the leading and trailing arm has decreased. We can thus expect the growth rate of the stream to decrease.

\subsubsection{A long stream in a rotating potential (Model D)}
In the fourth column of panels in Figure \ref{fig:2D_special_cases_Lz} we consider another simple stream model evolved in the presence of rotating bar which is now rotating slightly faster ($\Omega_{\rm b}=-61.3\kmskpc$, Model D). From the top panel we see that the orbit in the Galactic rest frame looks almost identical to the previous cases. However, the orientation of the bar near pericentre is now different. The progenitor's pericentre is near the major axis of the bar and receives almost no torque. The leading arm which arrived earlier receives a predominantly positive torque, leading to an increase in angular momentum and a decrease in energy. The trailing arm receives the opposite torque and increase its energy. This pattern of torques is opposite to that in Model C. As a result, the difference in energy between the leading and trailing arm has increased and we expect the growth rate of the stream to increase.

\subsection{Stream length and the orbital resonance}

\begin{figure}
\begin{center}
	\includegraphics[angle=0,width=0.95\columnwidth]{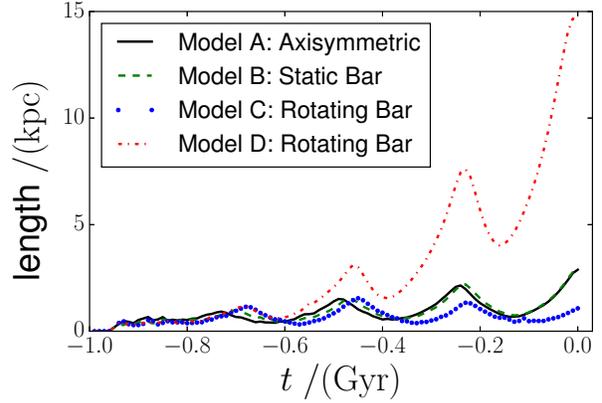}
\end{center}
\caption{
Stream length as a function of time
for the four representative stream models (Models A, B, C, and D) in Figure \ref{fig:2D_special_cases_Lz}.
Although these simulations end with the progenitor having the same location and velocity,
their time-evolutions are very diverse.
}
\label{fig:time_length_comparison}
\end{figure}

In Figure \ref{fig:time_length_comparison}, we show the evolution in the stream length for the four stream models in Figure \ref{fig:2D_special_cases_Lz}. As expected, we see that Model D which has pericentres near the major axis of the bar has the longest stream while Model C which has pericentres near the minor axis of the bar has the shortest stream. The static potential models (Model A and Model B) both give rise to a similar intermediate length stream since they cannot reshape the energy distribution of stream particles.

Since the location of the progenitor's pericentre controls whether the stream gets longer or shorter, there exists a resonance condition ($\Omega_{\rm b}=\Omega_\mathrm{res}$) to ensure that the progenitor approaches the bar in the same orientation for each pericentric passage, thus magnifying the effect. It is simplest to formulate this condition in the co-rotating frame since the bar is static in this frame. For each radial period, we require that the progenitor undergoes a half-integer number of angular periods in the co-rotating frame, i.e.
\eq{ \Omega_{\rm res} = \Omega_\phi \pm \frac{N}{2}\Omega_R , \label{eq:resonant_condition}}
where $\Omega_\phi$ is the angular frequency, $\Omega_R$ is the radial frequency, $N$ is an integer, and $\Omega_{\rm res}$ is the resonant frequency\footnote{We emphasise that in general these frequencies are not the epicyclic frequencies but rather describe the average rate of increase of the corresponding coordinates. In the limit of near circular orbits they reduce to the epicyclic frequencies. Thus this condition can be applied to orbits with significant eccentricities.}. This condition ensures that the progenitor approaches the bar in the same orientation modulo a rotation by 180$^\circ$ which leaves the bar's torque unchanged. Of course, this condition alone is not sufficient since the phase of the bar is also essential in determining how the growth rate will change. These resonant frequencies were used to select the bar pattern speeds in the examples shown in Figures \ref{fig:2D_special_cases_Lz} and \ref{fig:time_length_comparison}, resulting in extremely short and long streams.

We can now summarise the simple picture of bar shepherding which is encapsulated in Figure \ref{fig:2D_special_cases_Lz}. Put simply: particles in the stream approach the bar at different times and receive different torques and hence different changes to their energy. Depending on where along the bar the progenitor has a pericentre, this mechanism can either increase or decrease the stream's growth rate. If the progenitor has a pericentre near the major axis of the bar, the stream growth rate will increase. However, if the progenitor has a pericentre near the minor axis of the bar, the stream growth rate will decrease. We emphasise that the orbits in the Galactic rest frame in Figure \ref{fig:2D_special_cases_Lz} are almost identical but the presence of a rotating bar can lead to a dramatically different energy spread and hence stream length.

\section{Generating streams} \label{section:method}

In this section, we will describe the details of how we generate numerical models of stellar streams using the Lagrange point stripping method \citep{Kupper2012, Gibbons2014, Bowden2015}, as well as our potential choice and coordinate system. This method allows us to rapidly produce stellar streams and thus allows us to explore a large parameter space.

\subsection{Coordinate system}

Throughout this paper, we use two types of Galactocentric coordinate systems. The first one is the inertial coordinate system $(x,y,z)$
where the $(x,y)$-plane is the Galactic plane
and $z$-axis is directed toward the North Galactic Pole.
The current position of the Sun
is assumed to be
$(x, y, z)=(-R_0, 0, 0)$ with
$R_0 = 8.0 \;{\rm kpc}$.
The velocity of the Sun is assumed to be
$(v_{x}, v_{y}, v_{z})=(11.1, 244, 7.25)\;{\rm km\;s^{-1}}$
\citep{Schonrich2010,Bovy2012}, following \cite{Sesar2015}.

The second coordinate system is
the bar's co-rotating frame $(X_{\rm rot}, Y_{\rm rot}, z)$
in which the bar is aligned to the $X_{\rm rot}$-axis.
For later discussion,
we introduce the azimuthal angle
$\phi_{\rm rot} = \arctan(Y_{\rm rot}/X_{\rm rot})$
in this co-rotating frame. Finally,
we define $R$ 
to be the cylindrical Galactocentric radius which is the same in both systems.

\subsection{Models of the Milky Way} \label{section:potential}
\begin{figure}
\begin{center}
     \includegraphics[angle=0,width=0.95\columnwidth]{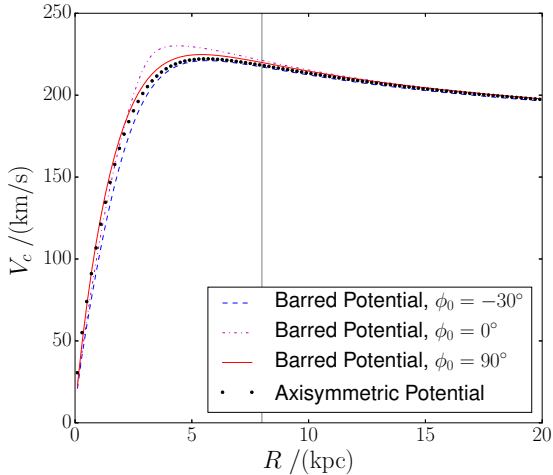}
\end{center}
\caption{
Rotation curves along the Sun-Galactic Centre line for our models.
For the barred potential,
three different orientations of the bar with respect to the Sun-Galactic Centre line are adopted.
The vertical line at $R=8 \kpc$ represents the Solar position.}
\label{fig:rotation_curve}
\end{figure}

We use two Milky Way potential models: an axisymmetric model and a barred potential. The rotation curves of these models are shown in Figure \ref{fig:rotation_curve}. The axisymmetric model is the three-component {\rm MWPotential2014} model \citep{Bovy2015} adopted in \cite{Sesar2015}, but with the bulge component replaced by a Hernquist profile \citep{Hernquist1990} with a mass of $5\times10^9\;M_\odot$ and a scale radius of $2 \kpc$.

Our barred potential is produced by replacing the Hernquist bulge in the axisymmetric model with the prolate bar model of \cite{Long1992}. The  symmetry axis lies in the disc $(x,y)$ plane, and the semi-major and semi-minor axes for the bar are taken to be
$a=3\;{\rm kpc}$ and $b=1\;{\rm kpc}$, respectively \citep{Bissantz2002}.
The pattern speed of the bar, $\Omega_{\rm b}$,
and the current bar inclination with respect to the Sun-Galactic centre line, $\phi_0$,
are free parameters. We define the sign of $\Omega_{\rm b}$ to be negative
if the bar's rotation is prograde.
This definition is mathematically natural in our coordinate system, since disc stars have negative values of $L_z$.
In Section \ref{section:2D}, we set $\phi_0=0 ^{\circ}$ for simplicity.
In Section \ref{section:3D}, we set $\phi_0=-30 ^{\circ}$ to match observations of the bar \citep{Lopez-Corredoira2005, Wegg2015},
where the negative sign means that the positive Galactic longitude $\ell$ part of the bar is closer to the Sun.

Figure \ref{fig:rotation_curve} shows the rotation curves of our axisymmetric potential and barred potential.
For the barred potential,
we show the rotation curves for different values of $\phi_0$. We see a slight dependence on $\phi_0$ of the rotation curve near
$R \simeq a = 3 \kpc$. However, the overall shape of the four rotation curves is very similar.

\subsection{Stream generation procedure}

We generate mock stellar streams
based on the Lagrange point stripping method
in the following manner:

\noindent (1) First we assume the position and velocity of the progenitor system
at the current epoch $t=0$ (`final condition').

\noindent (2) Then we rewind the progenitor's orbit backward in time from $t=0$ to $t=-\tau$
in the chosen potential.
\cite{Sesar2015} estimated the dynamical age of the Ophiuchus stream to be $\sim 300 \Myr$
by orbit fitting in an axisymmetric potential.
We fix $\tau = 1 \Gyr$ throughout this paper
to investigate if the dynamical age of the Ophiuchus stream
can be significantly older than $\sim 300 \Myr$ while keeping it as short as $\sim 2 \kpc$.

\noindent (3) Finally, we evolve the system from $t=-\tau$ to $t=0$
and release particles from two Lagrange points at a certain stripping rate.
The position and velocity of the progenitor at time $t$ are given by
$\vec{r}_{\rm prog}(t)$ and $\vec{v}_{\rm prog}(t)$, respectively.
The Galactocentric distances of the Lagrange points are given by
$r=r_{\rm prog}(t) \pm r_t(t)$, where $r_{\rm prog} \equiv |\vec{r}_{\rm prog}|$.
The tidal radius $r_t$ is defined by
\eq{
r_t = \left( \frac{G M_{\rm prog}}{{| \vec{r}_{\rm prog} \times \vec{v}_{\rm prog} |}^2 / r_{\rm prog}^4 - \partial^2 \Phi / \partial r_{\rm prog}^2}\right)^{1/3} ,
}
where $\Phi$ is the potential of the Milky Way.
The initial velocity of a stripped star with respect to the progenitor
is randomly drawn from an isotropic Gaussian distribution
with zero mean relative velocity.
The dispersion in each velocity component
is assumed to be $\sqrt{GM_{\rm prog}/(3 r_t)}$,
where $M_{\rm prog}$
is the progenitor's mass.
In this paper,
we assume that $M_{\rm prog}=10^4 \;M_\odot$ is constant in time.

\subsection{Definition of the length of the stream}

We define the length of the stream
as the arc length of the stream covering 80\% of stream stars.
We note that this length is different from the photometrically determined length
that is used in \cite{Sesar2015}. Instead, our definition can be thought of as a proxy for the length.

\section{Two-dimensional stream} \label{section:2D}

In Section \ref{section:theory} we gave a qualitative explanation of the effect of the bar on streams by inspecting three representative orbits. In this section, we will look at this effect quantitatively and in more detail by inspecting the evolution of the full distribution of stream particles. As in Section \ref{section:theory}, we will restrict the progenitor's orbit and the bar to lie in the same plane in order to give some physical insight. In Section \ref{section:3D}, we will consider the three dimensional problem.

\subsection{Lagrange-point stripping method simulation}

Here we consider a stellar stream generated from
a globular cluster of mass $10^4 \;M_\odot$
that has been disrupted for $\tau=1 \Gyr$.
We run a set of Lagrange-point stripping simulations
in either the axisymmetric potential or the barred potential.
For the barred potential,
we fix $\phi_0=0^{\circ}$ for simplicity and adopt a range of $\Omega_{\rm b}$.

We assume
that the progenitor is currently located at $(R,z)=(5, 0) \kpc$
with a prograde rotational velocity of $350 \kms$ and zero radial and vertical velocity.
In the simulations with a barred potential,
the progenitor's final location is
$\phi_{\rm rot}(t=0)=-30 ^{\circ}, 0 ^{\circ}$, or $90 ^{\circ}$ away from the bar.
In all the cases,
the progenitor experiences a pericentric passage at $t=0$
and moves within the disc plane.
We note that the angular velocity of the progenitor near the pericentre is
$\Omega_{\rm peri}= (-350 \kms) / (5 \kpc) = -70 \kmskpc$.

\subsection{Time evolution of the stream}
\begin{figure*}
\begin{center}
	\includegraphics[angle=0,width=0.475\columnwidth]{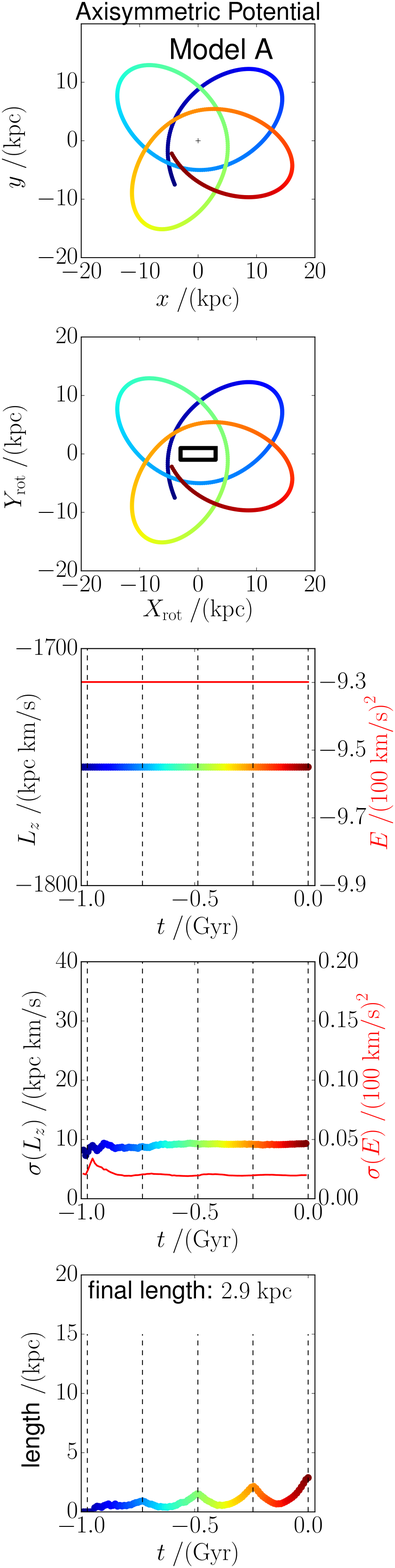}
	\includegraphics[angle=0,width=0.475\columnwidth]{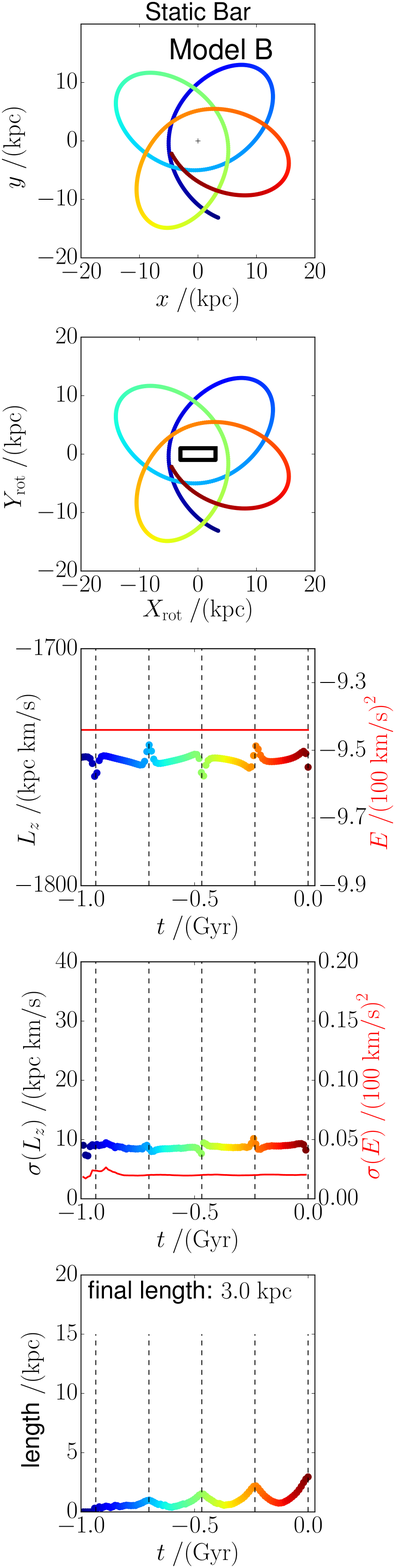}
	\includegraphics[angle=0,width=0.475\columnwidth]{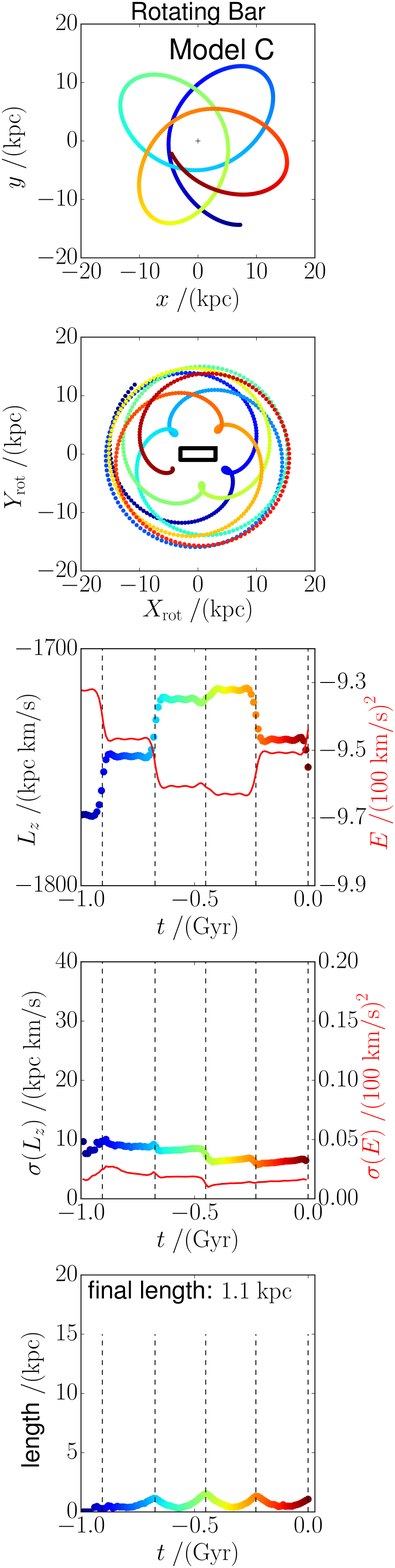}
	\includegraphics[angle=0,width=0.475\columnwidth]{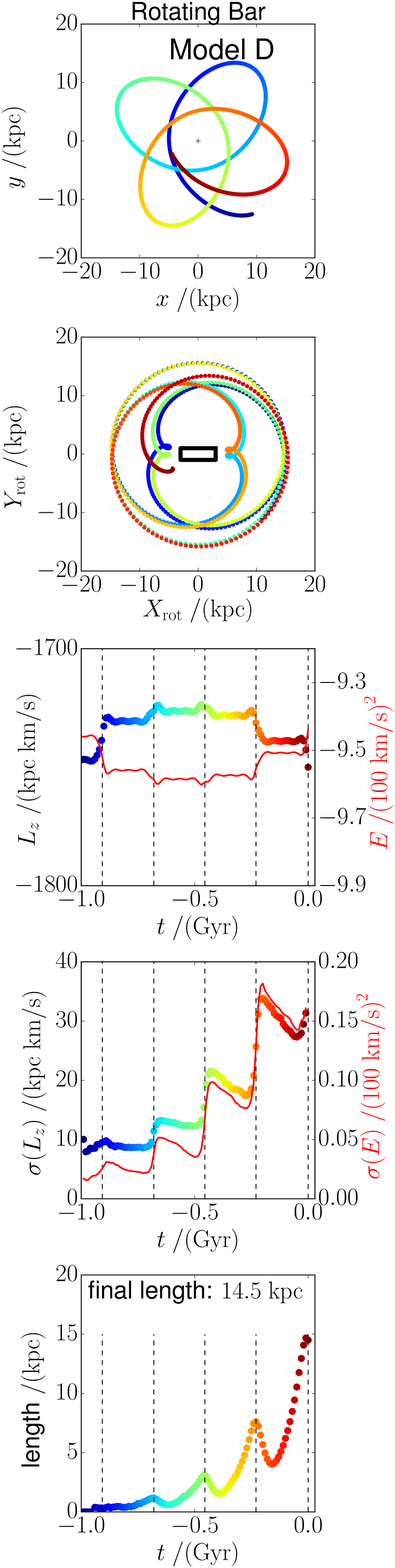}
\end{center}
\caption{
The time-evolution of the progenitor system and the associated stellar stream in our two-dimensional Lagrange-point stripping simulations.
In the four columns,
we show the results for four potential models (Models A, B, S, and L),
which are identical to the models in Figure \ref{fig:2D_special_cases_Lz}.
On the \textbf{top and second rows},
we show the progenitor's orbit in the Galactic rest frame and the co-rotating frame, respectively. The orbits are coloured by the the time with blue the earliest time ($t=-1\,\mathrm{Gyr}$) and red the latest ($t=0\,\mathrm{Gyr}$).
On the \textbf{third row}, we show the time-evolutions of $L_z$ (dots) and $E$ (red solid line) of the progenitor.
On the \textbf{fourth row},
we show the time-evolutions of dispersions $\sigma(L_z)$ (dots) and $\sigma(E)$ (red solid line) of the stream stars.
The \textbf{bottom row} shows the evolution of the stream length,
which are also presented in Figure \ref{fig:time_length_comparison}.
The vertical dashed lines in the third, fourth and the bottom rows show the times of pericentric passages of the progenitor.
}
\label{fig:2D_special_cases}
\end{figure*}

In order to demonstrate how a rotating bar affects the stream length,
here we focus on four representative simulations, Models A, B, C, and D,
for which the progenitor's final azimuthal angle is $\phi_{\rm rot}=-30 ^{\circ}$. These are the same four models as in Section \ref{section:theory}.
Model A is run in the axisymmetric potential,
and the other models are run in the barred potentials.
For Models B, C, and D, the pattern speeds are
$\Omega_{\rm b} / (\kmskpc) = 0, -58.1$ and $-61.3$, respectively.
The time-evolution of the progenitor and the stellar stream
for these models
are shown in Figure \ref{fig:2D_special_cases}.
These four models are identical to the four simulations shown in Figure \ref{fig:2D_special_cases_Lz}.

In Figure \ref{fig:2D_special_cases} we show the orbit of the progenitor in the inertial and co-rotating frame for our four models. We also show the evolution of $E$ and $L_z$ for the progenitor, $\sigma(L_z)$ and $\sigma(E)$ for the stream, and the stream length. This figure is a counterpart to Figure \ref{fig:2D_special_cases_Lz} where we looked at the progenitor and two representative stars in the leading and trail arms. In Figure \ref{fig:2D_special_cases} we see that the progenitor's orbits in the inertial frame are almost identical for these models, reflecting the fact that these potentials have similar rotation curves and that the progenitors' final conditions are identical. However, they look very different in the co-rotating frame. This difference causes different evolutions of the stream length as we argued in Section \ref{section:theory}. We will now examine how the energy and angular momentum, as well as their dispersions and the stream length, are affected by the bar.

\subsubsection{Variation of the progenitor's $(E, L_z)$}

Essentially, the role of the bar on the stellar stream
is to change the energy and angular momentum of the progenitor and stream stars.
Thus, understanding the time-evolution of $(E,L_z)$ of the progenitor
is an important step in our study.

In our barred potentials,
the angular momentum of the progenitor, $L_z$,
is not conserved.
The change in $L_z$ is most prominent near pericentre,
where the non-axisymmetry due to the bar is most important.
Since the potential is almost axisymmetric at large $R$,
$L_z$ is nearly conserved between consecutive pericentric passages.

Figure \ref{fig:torque} shows the contour plot of the torque from the bar,
which is responsible for the change in $L_z$.
We see that the torque vanishes on
both the $X_{\rm rot}$- and $Y_{\rm rot}$-axes.
This property means that $L_z$ does not change notably
when the progenitor's pericentre
is located near the major axis of the bar
or near the minor axis.
We also see from Figure  \ref{fig:torque} that
$L_z$ increases (positive torque)
in the first and third quadrants of the $(X_{\rm rot}, Y_{\rm rot})$-plane
and that $L_z$ decreases (negative torque)
in the second and fourth quadrants.
This is why the progenitor's $L_z$ shows a sharp rise or drop
when the pericentre location is far from both the $X_{\rm rot}$- and $Y_{\rm rot}$-axes. In the rotating barred potentials,
the changes in $E$ and $L_z$ are related since the Jacobi energy is conserved [see equation (\ref{eq:jacobi})].

\subsubsection{Variation in the dispersions $\sigma(L_z)$ and $\sigma(E)$ of stream stars}

Since the dispersions, $\sigma(L_z)$ and $\sigma(E)$, of stream stars
govern the growth rate of the stream,
it is important to understand
how these quantities are affected by the rotating bar.

As we can see from the fourth row of Figure \ref{fig:2D_special_cases}, stream stars in our four models
exhibit different evolutions of the dispersions $\sigma(E)$ and $\sigma(L_z)$. This difference can be understood using Figure \ref{fig:2D_special_cases_Lz} in which we presented the evolution of three representative orbits. These three representative orbits can be considered as samples from the full distributions shown in Figure \ref{fig:2D_special_cases}. In Figure \ref{fig:2D_special_cases_Lz},
we showed the orbital evolutions of the progenitor system (black solid line),
a typical leading-tail star (blue dotted line),
and a typical trailing-tail star (red dashed line)
from $t=-0.55 \Gyr$ to $t=-0.35 \Gyr$.

In Model A,
both $L_z$ and $E$ are conserved,
so that the differences in $L_z$ and $E$
between the leading and trailing stars
do not change during this pericentric passage. Correspondingly $\sigma(E)$ and $\sigma(L_z)$ in Figure \ref{fig:2D_special_cases} do not evolve significantly over the course of the simulation.

The situation is similar in Model B.
In this case,
the three orbits in the co-rotating frame look almost identical,
so the three orbits feel the torque from the bar in a similar fashion.
As a result,
the difference in $L_z$ between the leading and trailing stars
is almost conserved.
The difference in $E$
is also conserved since each orbit conserves $E$. As with Model A, the $\sigma(E)$ and $\sigma(L_z)$ of Model B from Figure \ref{fig:2D_special_cases} do not evolve significantly over the course of the simulation.

In contrast, the situation in the rotation barred potentials (Models C and D), is rather different. In Model C,
the three orbits in the co-rotating frame look notably different from each other.
Before the pericentric passage,
$L_z$ is larger for the leading star than for the trailing star.
At the pericentre of the orbit,
the leading and trailing stars receive negative and positive torques from the bar, respectively,
reflecting their pericentre locations in the co-rotating frame. Inspecting Figure \ref{fig:2D_special_cases} we note that this behaviour corresponds to a small decreases in $\sigma(L_z)$, most notably around the third pericentric passage at $t\approx-0.45\Gyr$.
As a result,
after this pericentric passage,
the gap in $L_z$ between the leading and trailing stars shrinks;
and the gap in $E$ also shrinks.
In Model D,
on the other hand,
the gaps in $L_z$ and $E$ increase,
since the leading and trailing stars receive positive and negative torques, respectively. Figure \ref{fig:2D_special_cases} shows that in this case $\sigma(E)$ and $\sigma(L_z)$ increase significantly over the course of the simulation. The increases in the dispersions occur around pericentric passage and, as the simulation assumes a constant stripping rate, between pericentric passage the dispersions decay due to new material entering the stream.

\subsubsection{Variation in the stream length}

Here we investigate how the
changes in $\sigma(E)$ of stream stars are translated into the
time-evolution of the stream length. On the bottom row in Figure \ref{fig:2D_special_cases},
we show the time-evolution of the stream length.
These lengths are identical in Figure \ref{fig:time_length_comparison}.
In all of these cases,
the stream length shows an oscillatory behaviour
as it stretches and compresses near pericentres and apocentres respectively.
In Models A and B,
$\sigma(E)$ is nearly constant in time,
so that the stream gradually grows.
In Model C,
we see a notable decrease in $\sigma(E)$ (especially near the third pericentric passage),
which suppresses the growth rate of the stream and results in a short stream.
In Model D, on the other hand,
we see a notable increase in $\sigma(E)$,
which enhances the growth rate and results in a long stream.
From these representative models,
we can see how a rotating bar affects the length of a stream.

\subsection{Final length of the stream as a function of $\Omega_{\rm b}$}

\begin{figure}
\begin{center}
	\includegraphics[angle=0,width=0.95\columnwidth]{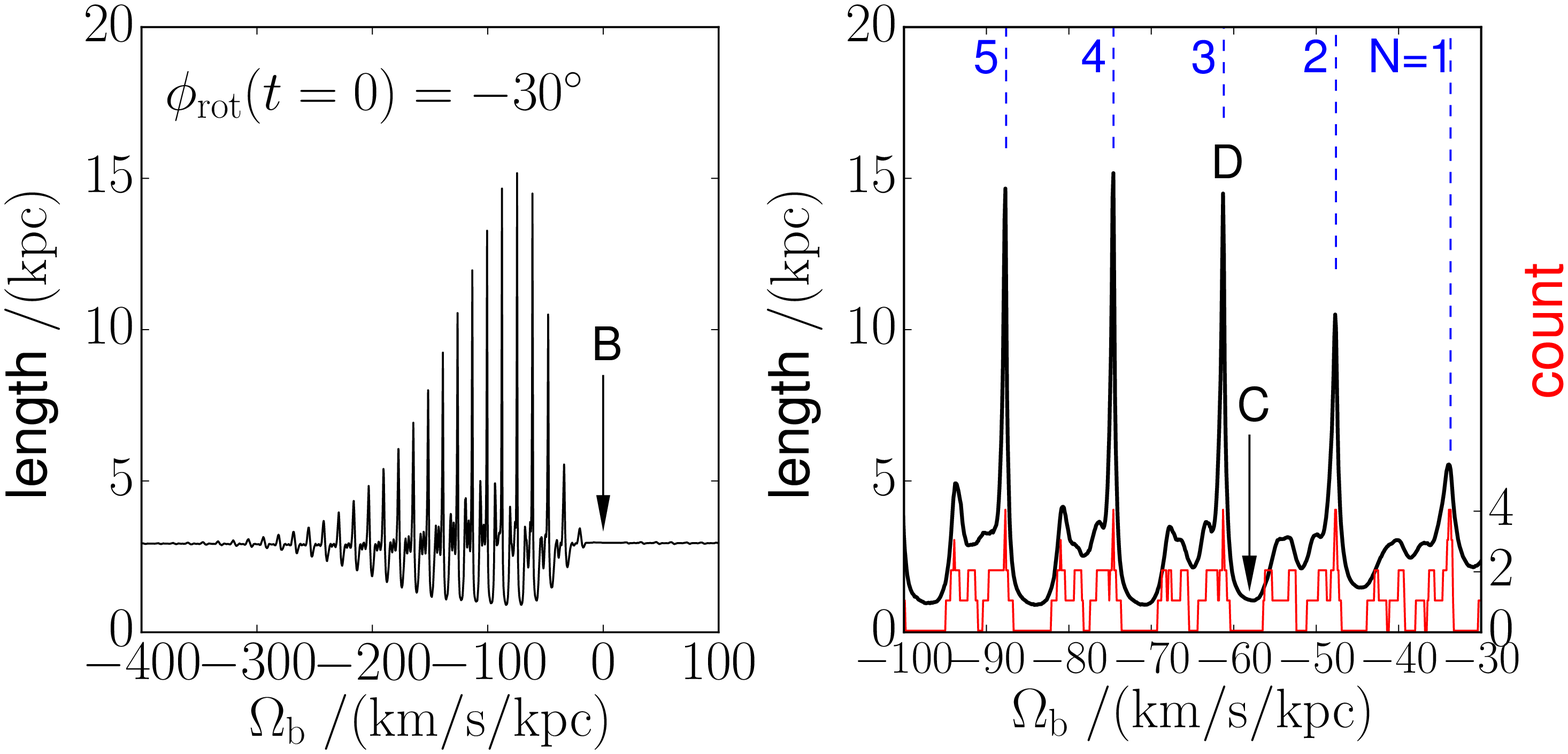}\\
	\includegraphics[angle=0,width=0.95\columnwidth]{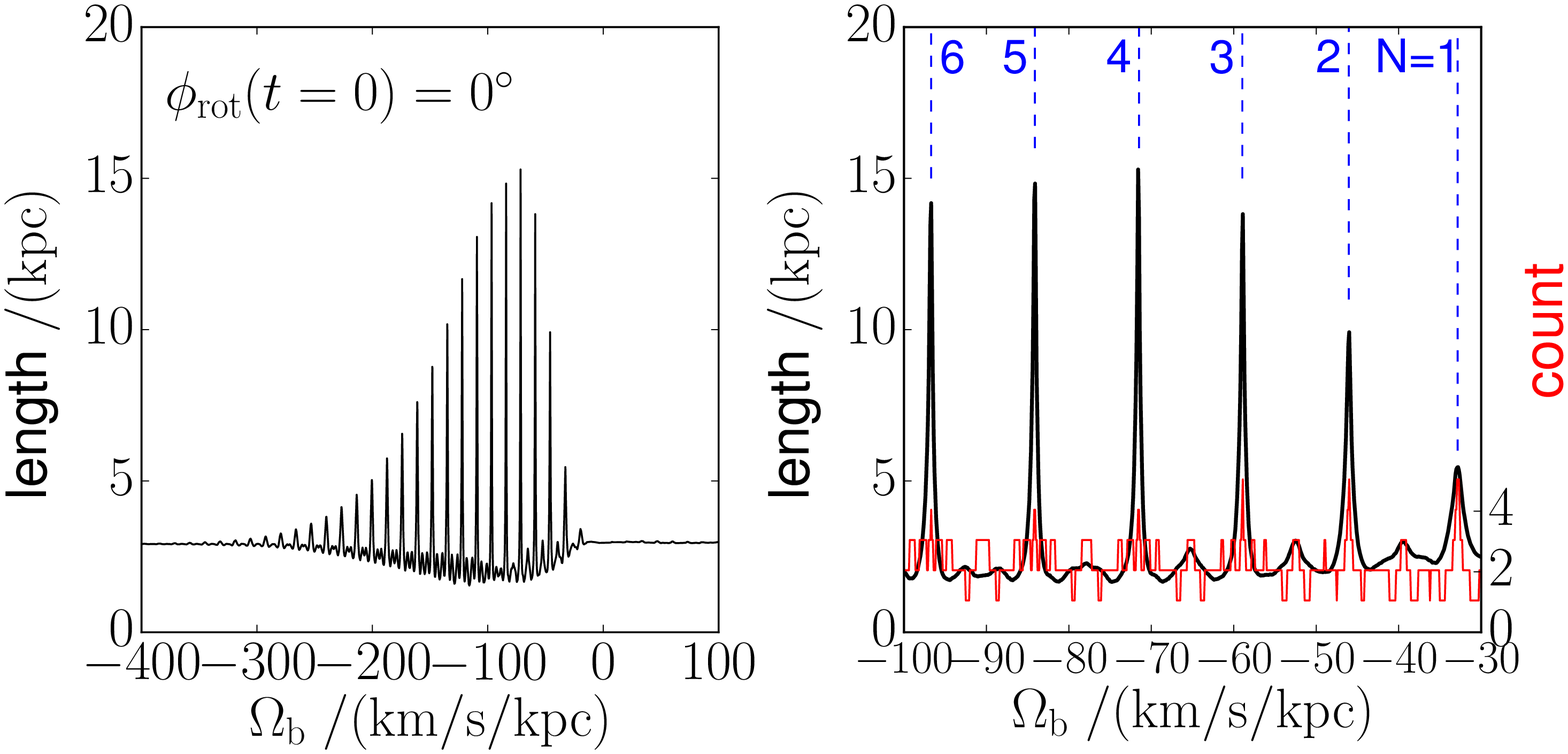}\\
	\includegraphics[angle=0,width=0.95\columnwidth]{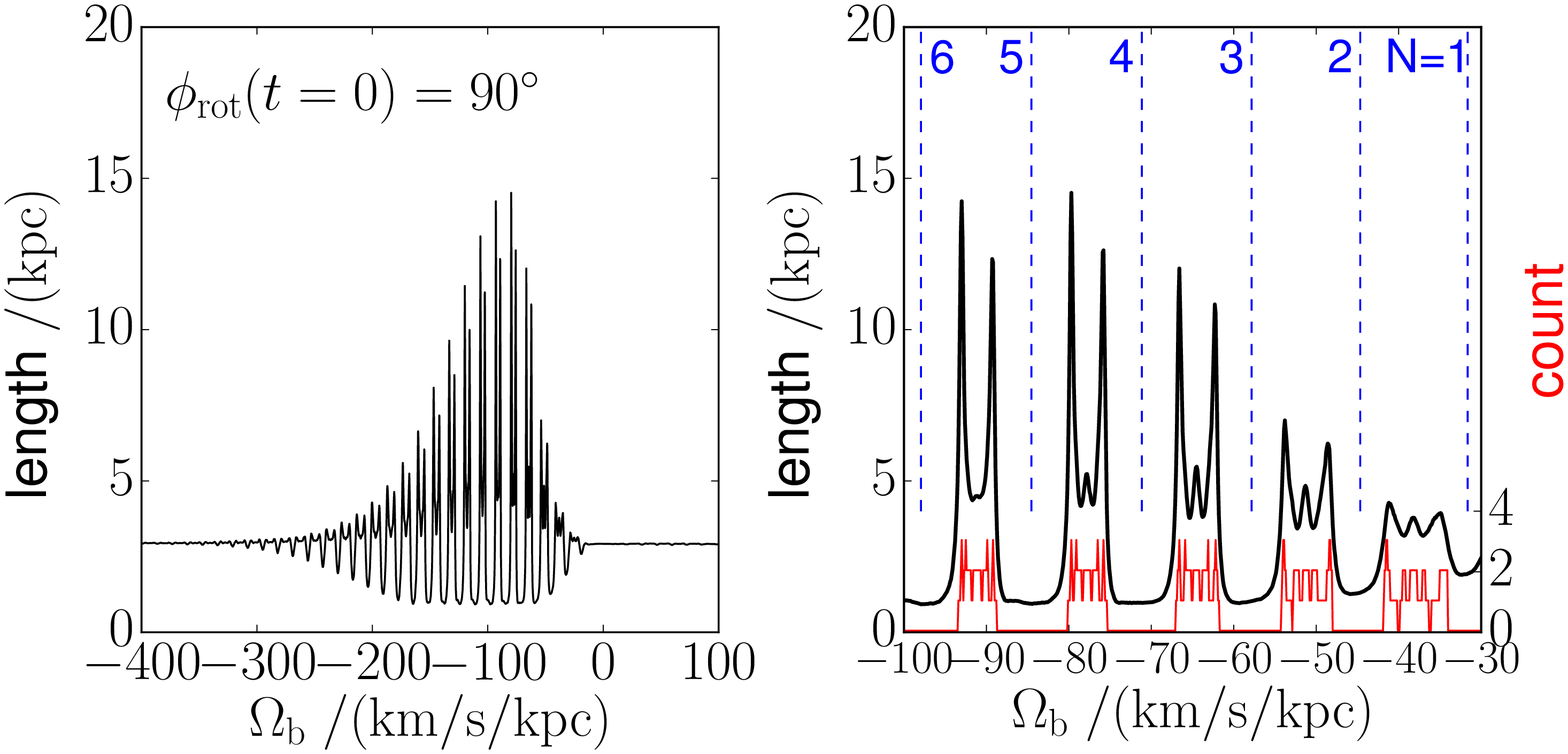}
\end{center}
\caption{
The final stream length as a function $\Omega_{\rm b}$ for our two-dimensional stream simulations in barred potentials.
From top panel to bottom,
we show the results for different final conditions
(azimuthal angle of the progenitor in the co-rotating frame $\phi_{\rm rot}(t=0)= -30 ^{\circ}, 0 ^{\circ}$ and $90 ^{\circ}$).
The left-hand panels show the results for $-400 \kmskpc \leq \Omega_{\rm b} \leq 0 \kmskpc$,
and the right-hand panels show the results for $-100 \kmskpc \leq \Omega_{\rm b} \leq -30 \kmskpc$.
The red thin line shows the number of
pericentric passages at which the progenitor passes near the major axis of the bar.
The dashed vertical lines shows the locations of the resonance frequencies $\Omega_{\rm res}$ in equation (\ref{eq:resonant_condition})
for several values of $N$.
Models B, C, and D correspond to
$\Omega_{\rm b}/(\kmskpc)=0, -58.1$ and $-61.3$, respectively,
with $\phi_{\rm rot}(t=0)= -30 ^{\circ}$;
and they are marked on the top panel.
Note that negative values of $\Omega_{\rm b}$ correspond to bars with prograde rotation.
}
\label{fig:2D_L_vs_Omega}
\end{figure}

Now that we understand the basic mechanism, we can examine how varying the pattern speed changes the stream length. In Figure \ref{fig:2D_L_vs_Omega} we show
the final stream length as a function of $\Omega_{\rm b}$ for different final progenitor positions, $\phi_{\rm rot}(t=0)$. For all values of $\phi_{\rm rot}(t=0)$,
the final length of the stream converges to an asymptotic value of $2.9$ kpc
as $|\Omega_{\rm b}| \to \infty$.
This is because, in this limit,
the potential changes so rapidly that
it becomes effectively axisymmetric.
Indeed,
when the axisymmetric potential is adopted,
the final stream length is $2.9$ kpc,
which is identical to the asymptotic length.

Figure \ref{fig:2D_L_vs_Omega} shows that
the final length is a highly oscillating function of $\Omega_{\rm b}$.
The amplitude of this oscillation is maximal
at around $\Omega_{\rm b} \simeq \Omega_{\rm peri} = -70 \kms$,
since the interaction between the stream stars and the bar is most efficient
when their angular velocities are similar.
At $\Omega_{\rm b}>0$ (when the bar and the progenitor rotate in the opposite direction),
we see a small oscillation of the final length as a function of $\Omega_{\rm b}$. However, this is not as significant as at $\Omega_{\rm b}<0$
since the progenitor's angular velocity at pericentre is large in the co-rotating frame. As a result, each star in the stream receives a similar change in angular momentum and hence there is no change in the stream growth rate.

In Section \ref{section:theory},
we showed that the width of the energy distribution of stream stars
increases when the progenitor's pericentre is located near the major axis of the bar
and decreases when it is located near the minor axis.
This finding naturally indicates that
the stream length is maximised and minimised
when the progenitor's pericentres are always near the major axis of the bar
and always near the minor axis, respectively.
This idea can be tested with our simulations.

For $-100<\Omega_{\rm b}/(\kmskpc)<-30$,
the progenitor system experiences $4$-$5$ pericentric passages.
For each value of $\Omega_{\rm b}$,
we count the number of  pericentric passages
at which the progenitor's azimuthal angle $\phi_{\rm rot}$ is smaller than $20 ^{\circ}$.
We see
from Figure \ref{fig:2D_L_vs_Omega}
that the strong peaks and troughs in the final length
correspond, respectively, to the peaks and troughs of the `count'.
This correspondence nicely demonstrates that
the final length of the stream is elongated or shortened
if the progenitor passes at a similar orientation to the bar at all of its pericentres
[as described by the resonant condition (\ref{eq:resonant_condition})].

In addition to the resonant frequency given by equation (\ref{eq:resonant_condition}), the stream length in Figure \ref{fig:2D_L_vs_Omega} has features which look like higher order resonances. For example, in the top panel we see small peaks between the larger peaks. We can generalise our resonant frequency by requiring that for every $K$ radial periods, the progenitor undergoes $N/2$ azimuthal periods in the co-rotating frame. This gives the generalised resonance frequency
\eq{ \Omega_{\rm res,general} = \Omega_\phi \pm \frac{N}{2K} \Omega_R . \label{eq:generalized_resonance} }
This form of resonance frequencies has been studied in the context of the orbital resonance of disc stars in rotating spiral potentials
\citep{Contopoulos1970, Lynden-Bell1972, Shu1992}, but has not been studied in the context of stellar streams.
We note that we do not limit our discussion to nearly circular orbits, which is in contrast with Galactic disc dynamics.
For $K=1$, the generalised resonance frequency reproduces our resonant frequency in equation (\ref{eq:resonant_condition}). For $K=2$, we would expect features with half of the frequency spacing of the main resonance. The effect on the stream length at these frequencies should not be as strong as the main resonance since the bar is only approached in the same orientation every $K$ radial oscillations. As a result, the bar's effect will not be as coherent. This naturally explains the small peaks between large peaks in the top two panels of Figure \ref{fig:2D_L_vs_Omega}. Higher order features with $K=3$ also appear to be present in the bottom panel of Figure \ref{fig:2D_L_vs_Omega}. We will explore this higher order resonance in future work.

For $\phi_{\rm rot}(t=0)=0 ^{\circ}$ case,
the progenitor's final pericentric passage
is located on the major axis of the bar.
Thus,
if the condition of $\Omega_{\rm b} =\Omega_{\rm res}$ is satisfied,
the progenitor's pericentres are always located near the major axis,
generating a long stream.
For $\phi_{\rm rot}(t=0)=90 ^{\circ}$ case,
on the other hand,
the same condition guarantees the progenitor
to pass near the minor axis at every pericentric passage,
generating a short stream.
Indeed,
as seen in Figure \ref{fig:2D_L_vs_Omega},
the positions of the resonance frequencies $\Omega_{\rm res}$
(calculated by using the algorithm in \citealt{SandersBinney2014}) nicely match
the positions of the peaks for the cases of $\phi_{\rm rot}(t=0)=-30 ^{\circ}$ and $0 ^{\circ}$;
and the positions of the troughs for the case of $\phi_{\rm rot}(t=0)= 90 ^{\circ}$.

\section{Application to the Ophiuchus stream} \label{section:3D}

In Sections \ref{section:theory} and \ref{section:2D} we explored how a rotating bar can re-shape the energy distribution of a stream, and hence its growth rate. This analysis was restricted to 2D for simplicity which allowed us to build an intuitive picture. Now, we will study the effect in 3D by considering the effect of the bar on the Ophiuchus stream. We will discuss the case for arbitrary orbital geometries and pattern speeds in a separate paper.

\subsection{Numerical simulations of Ophiuchus-like streams}

\subsubsection{Lagrange-point stripping simulations}\label{section:Lagrange3D}

Here we describe how we
generate mock Ophiuchus-like streams with our Lagrange-point stripping method. We assume that
the progenitor system of the Ophiuchus stream is currently located at
$(\ell, b)=(5.0, 31.37) ^{\circ}$ and that it has a distance modulus of $DM=14.57$
and a heliocentric line-of-sight velocity of $v_{\rm los}=289.1 \kms$.
The 3D position and the line-of-sight velocity
correspond to the central point of the best-fit model of the Ophiuchus stream in \cite{Sesar2015}.
We also assume that the proper motion of the progenitor is
$(\mu_{\ell*}, \mu_b)=(-7.7, 1.5) {\rm mas \; yr^{-1}}$,
following the best-fit orbit from \cite{Sesar2015}.
We note that our choice of proper motion
guarantees a rough alignment of the progenitor's velocity vector and the stream's direction.
As discussed in \cite{Sesar2015},
these values of proper motion deviate from the purely observationally determined proper motion.
We assume the same $R_0$,
LSR velocity and the Solar peculiar motion as in \cite{Sesar2015}.
Under these conditions,
the progenitor's current position is
$(x, y, z)=(-1.0223, 0.61047, 4.270444) \kpc$
and its current velocity is
$(v_x, v_y, v_z) = (252.848807, -35.435133, 207.550556) \kms$.

We assume that the progenitor of the Ophiuchus stream was
a relatively light cluster with mass of $10^4 M_\odot$,
which is close to the lower limit ($0.7 \times 10^4 M_\odot$)
given in \cite{Sesar2015} and identical to the mass used in their $N$-body model.

In reality, the stripping of stars from the progenitor is
most prominent near the pericentre (e.g., \citealt{Dehnen2004}).
In order to model the stripping realistically,
we assume the stripping rate to be a sum of Gaussians
peaked at the time of pericentric passages with a dispersion of $10 \Myr$.

The observationally determined values of $\Omega_{\rm b}$ of the Galactic bar are distributed around
$-70 \leq \Omega_{\rm b} / (\kmskpc) \leq -20$
\citep{Binney1997, Dehnen1999, Debattista2002, Rodriguez-Fernandez2008, Antoja2014}.
Here we adopt a rather wide range of $\Omega_{\rm b}$ of the bar
of $-100 \leq \Omega_{\rm b} / (\kmskpc) \leq 0$.

\subsubsection{$N$-body simulations}\label{section:Nbody}
To validate our Lagrange-point stripping simulations,
we have run several $N$-body simulations of a King profile cluster disrupting in potentials with various pattern speeds, $\Omega_{\rm b}$. These simulations are run with the $N$-body part of \textsc{Gadget-3}, which is similar to \textsc{Gadget-2} \citep{Springel2005}. We have modified this code by including the effect of static and rotating potentials. We model the clusters as King profiles with $M=10^4 M_\odot$, $W=2$, and $r_{\rm core} = 30\pc$. These parameters are identical to those used in the $N$-body simulations in \cite{Sesar2015} so they can be directly compared. Each cluster is represented by $10^5$ particles and a softening of $1 \pc$ was used. We compare our $N$-body simulations against the Lagrange-point stripping method in Section \ref{section:final_length} and against observations of the Ophiuchus stream in Section \ref{section:observations}.

\subsection{Time evolution of the stream} \label{section:3D_time_evolution}

Next, we investigate how Ophiuchus-like streams evolve. In Figure \ref{fig:3D_special_cases}
we show the time evolution of the progenitor and the stream stars
for two representative runs of the Lagrange-point stripping simulations
with $\Omega_{\rm b}/(\kmskpc)=-50.8$ and $-56.6$,
which result in a long and a short stream, respectively.
We see that the changes in $(E,L_z)$ are prominent only when
the progenitor experiences its pericentric passages.
This property is reminiscent of our 2D simulations (Figure \ref{fig:2D_special_cases}),
where these changes are notable only at the pericentres of the progenitor.

Figure \ref{fig:3D_special_cases} also shows that
$\sigma(L_z)$ and $\sigma(E)$
have a notable rise or drop only at the pericentric passages,
just as in our 2D simulations. However, unlike the 2D simulations that used a constant stripping rate, the bursty stripping at pericentric passage leads to no decay in the dispersions between pericentres.
In addition, we see that
the decrease in $\sigma(E)$ causes a suppressed growth of the length
for $\Omega_{\rm b} = -56.6 \kmskpc$ simulation
and
the increase in $\sigma(E)$ causes a rapid growth of the  length
for $\Omega_{\rm b} = -50.8 \kmskpc$ simulation.

\subsubsection{Shepherding mechanism of the bar} \label{section:3D_three_orbits}

\begin{figure*}
\begin{center}
	\includegraphics[angle=0,width=0.8\columnwidth]{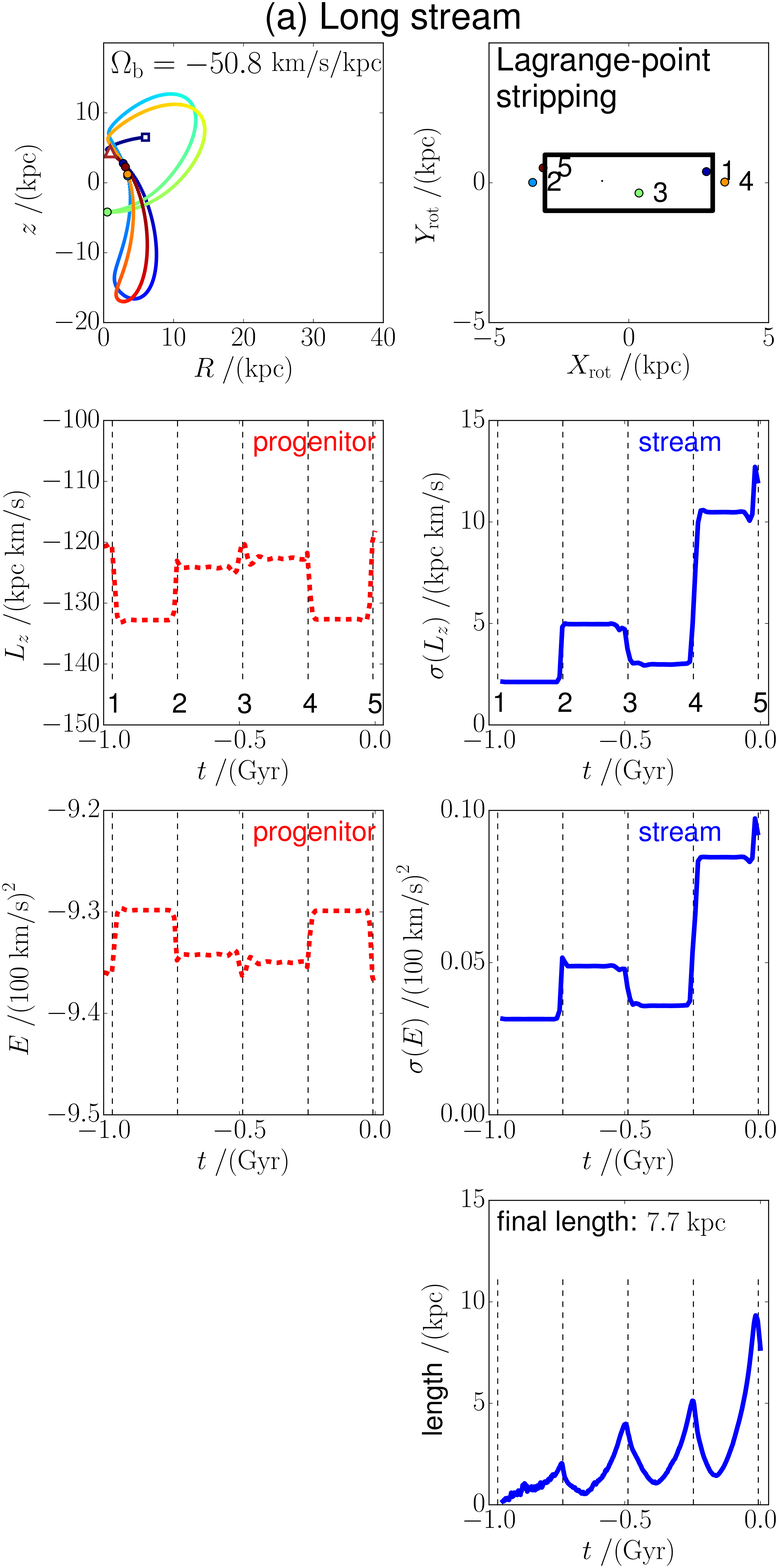}
	\includegraphics[angle=0,width=0.8\columnwidth]{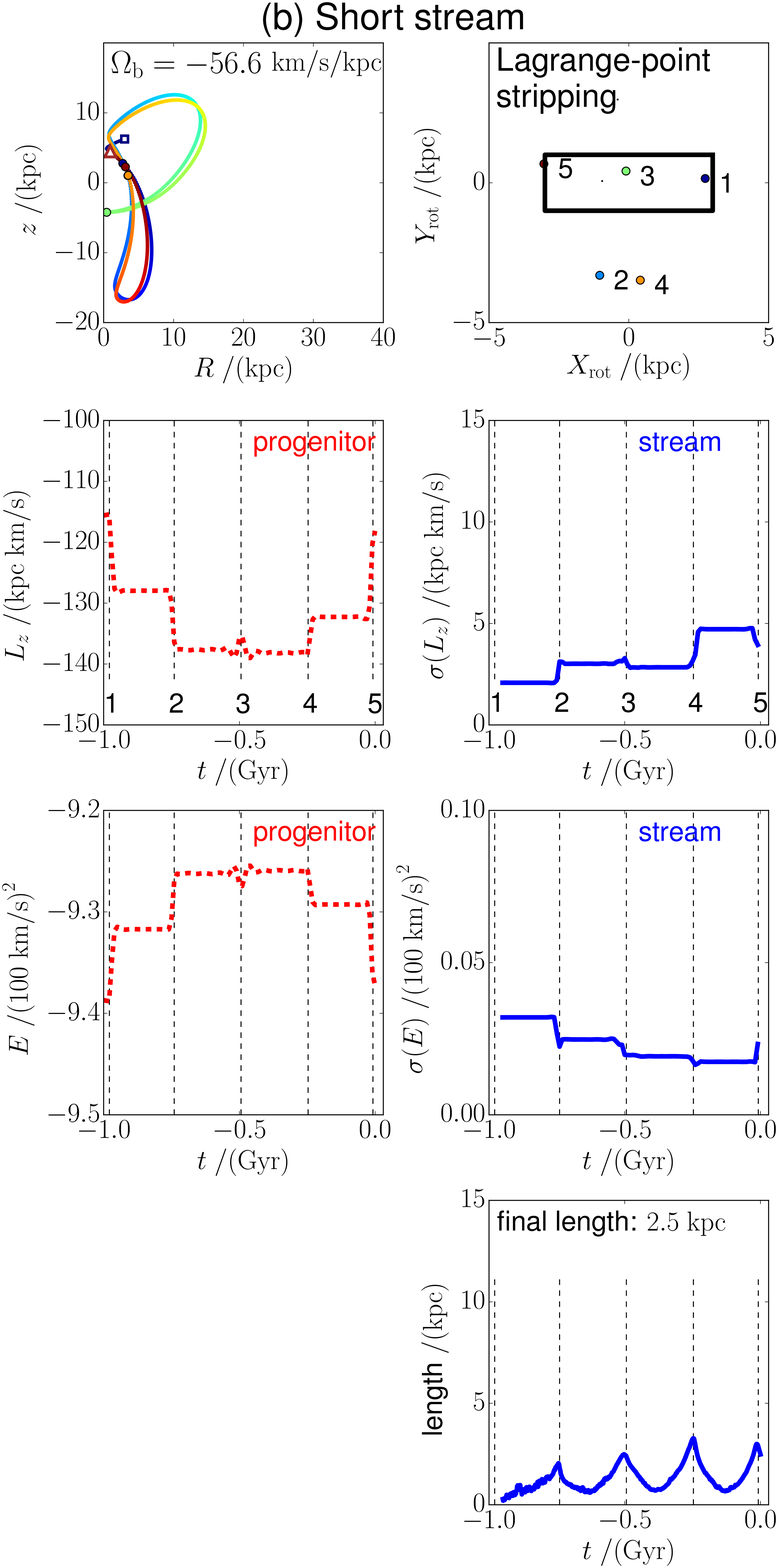}
\end{center}
\caption{
The time-evolution of the progenitor system and the stellar stream
generated in our three-dimensional Lagrange-point stripping simulations.
The left-hand two columns [marked by \textbf{(a)}] and the right-hand two columns [marked by \textbf{(b)}]
show the results for
$\Omega_{\rm b}/(\kmskpc) = -50.8$ and $-56.6$ (prograde bar), respectively.
On the top panels,
we show the orbital properties of the progenitor system. The orbits are coloured by the the time with blue the earliest time ($t=-1\,\mathrm{Gyr}$) and red the latest ($t=0\,\mathrm{Gyr}$).
The first panel on the top row shows the orbits in the $(R,z)$-plane.
The progenitor's positions at $t=-1 \Gyr$ and $t=0 \Gyr$ are shown
with an open blue square and an open red triangle respectively.
Also, the positions of the progenitor's pericentres are shown with circles.
The second panel on the top row shows the positions of the progenitor's pericentres
projected onto the co-rotating $(X_{\rm rot}, Y_{\rm rot})$-plane of the bar.
The number $i$ ($1 \leq i \leq5$) represents the $i$th pericentric passage.
On the second and third rows,
we show the time-evolutions of the progenitor's $(L_z, E)$ (red dashed lines) and
the time-evolutions of the dispersions $(\sigma(L_z), \sigma(E))$ of the stream stars (blue solid lines).
On the bottom row,
we show the time-evolution of th stream length.
The vertical dashed lines in the second, third, and the bottom rows
show the times of pericentric passages of the progenitor.
}
\label{fig:3D_special_cases}
\end{figure*}

Here we have a closer look at the second pericentric passage
in the $\Omega_{\rm b} = -56.6 \kmskpc$ simulation
to demonstrate the shepherding mechanism of the bar.
We choose this pericentric passage
because $\sigma(E)$ decreases most prominently in Figure \ref{fig:3D_special_cases}; and
because we see a counter-intuitive behaviour that $\sigma(E)$ decreases while increasing $\sigma(L_z)$.

\begin{figure*}
\begin{center}
	\includegraphics[angle=0,width=0.8\textwidth]{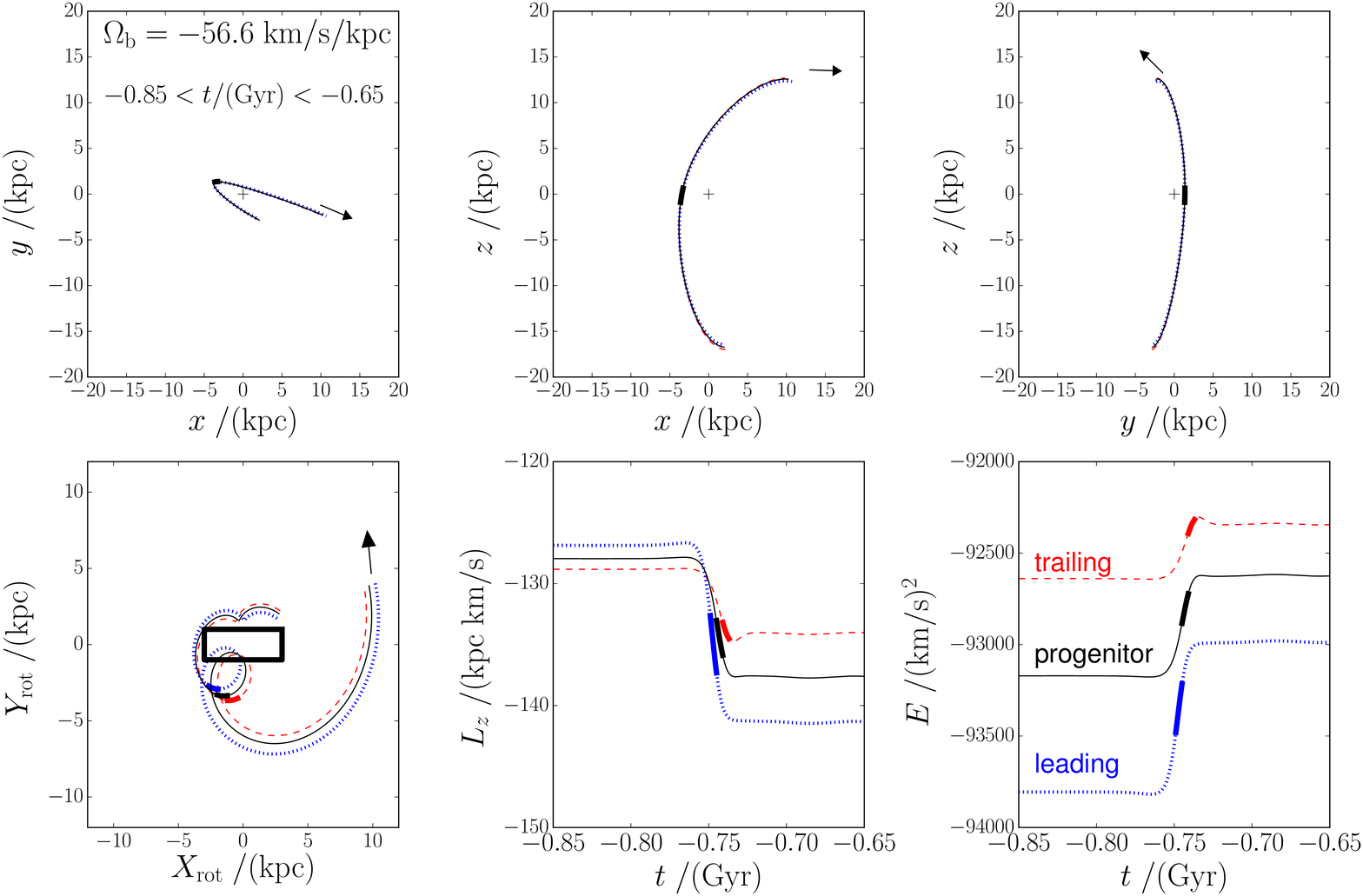}
\end{center}
\caption{
The time-evolution of the progenitor system (black solid line), a leading-tail star (blue dotted line), and
a trailing-tail star (red dashed line)
near one of the pericentric passages in the Lagrange-point stripping simulation with $\Omega_{\rm b}=-56.6 \kmskpc$  (prograde bar)
that reproduce an Ophiuchus-like short stream (see Section \ref{section:3D}).
On the top panels, we show the orbits in three projections of the Galactic rest frame.
On the bottom-left panel, we show the orbits in the co-rotating frame.
On the middle and right panels on the bottom row,
we show the time-evolution of $L_z$ and $E$, respectively.
The thickened part of the lines corresponds to the disc-crossing ($|z|<1 \kpc$).
}
\label{fig:3D_special_cases_track}
\end{figure*}

In Figure \ref{fig:3D_special_cases_track}
we show the orbit of the progenitor, a leading-tail star, and a trailing-tail star
as they cross the disc plane
and then (only $\sim 3 \Myr$ later) experience a pericentric passage.
These orbits cross the disc plane at a similar position in the $(x,y)$-plane,
but at different positions in the $(X_{\rm rot}, Y_{\rm rot})$-plane,
due to the different disc-crossing times.
As a result,
these orbits receive different torques from the bar.
Before the disc crossing (or the pericentric passage),
the leading star has the largest value of $L_z$.
However,
since the leading star has the greatest loss in $L_z$,
$(L_{z, {\rm leading}} - L_{z, {\rm trailing}})$ becomes negative
and $| L_{z, {\rm leading}} - L_{z, {\rm trailing}} |$ increases.
This is why $\sigma(L_z)$ increases.
On the other hand,
the gap in $E$ between the leading and trailing stars shrinks,
in a similar fashion as in Model C in Figure \ref{fig:2D_special_cases_Lz}.
The full explanation is due to the underlying distribution in $E$ and $L_z$ of the debris and is given in the Appendix.

\subsection{Final length as a function of $\Omega_{\rm b}$} \label{section:final_length}

\begin{figure}
\begin{center}
	\includegraphics[angle=0,width=0.95\columnwidth]{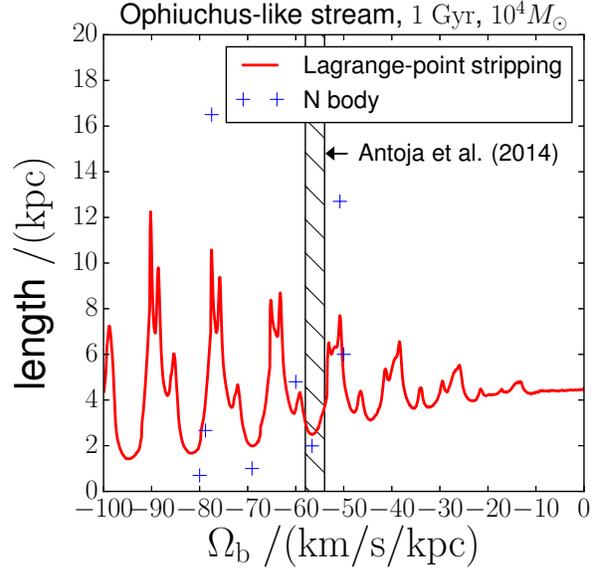}
\end{center}
\caption{
The final stream length as a function $\Omega_{\rm b}$ for our Ophiuchus-like stream simulations in barred potentials.
The red solid line shows the results from our Lagrange-point stripping simulations,
and the blue crosses show the results from our $N$-body simulations.
The vertical band shows the recent constraint on the Galactic bar's pattern speed of
$\Omega_{\rm b}=-56 \pm 2 \kmskpc$ \protect\citep{Antoja2014}.
Note that negative values of $\Omega_{\rm b}$ correspond to bars with prograde rotation.
}
\label{fig:3D_length_Omega}
\end{figure}

Figure \ref{fig:3D_length_Omega}
shows the relationship between the final stream length and $\Omega_{\rm b}$
for our simulations. We see that the final stream length of the Lagrange-point stripping simulations
shows a periodic pattern as a function of $\Omega_{\rm b}$,
very similar to the 2D case in Figure \ref{fig:2D_L_vs_Omega}.
At $\Omega_{\rm b}/(\kmskpc)\simeq -82, -69$ and $-56$,
we see troughs in the final length.
These troughs suggest that
the Ophiuchus stream can be as short as $\sim 2 \kpc$
even if its dynamical age is $\simeq 1 \Gyr$.
Also, we note that the final length changes by a factor of 5 or more for a modest change in $\Omega_{\rm b}$.

The results from the Lagrange-point stripping simulations
are justified by the more realistic $N$-body results,
since they are broadly consistent.
For example,
in both types of simulations,
models with $\Omega_{\rm b}/(\kmskpc) = -77.5$ and $-50.8$
result in long streams,
and models with  $\Omega_{\rm b}/(\kmskpc) = -80.0, -69.1$ and $-56.6$
result in short streams.
Also,
the length of the stream is sensitive to $\Omega_{\rm b}$ at $-80 < \Omega_{\rm b}/(\kmskpc) < -77.5$
in both cases.

We note that
the $N$-body simulations tend to result in streams with more extreme lengths.
For example,
the final lengths of the simulations with $\Omega_{\rm b}/(\kmskpc) = -69.1$ and $-77.5$
are $2.0 \kpc$ and $10.6 \kpc$ for the Lagrange-point stripping models;
but are $1.0 \kpc$ and $16.5 \kpc$ for the $N$-body models.
However, our Lagrange-point stripping method is useful
to search for conditions for short streams
since it can generate a model in $\sim 3$ seconds.

\subsubsection{Resonance frequencies}

In Figure \ref{fig:3D_length_Omega},
the intervals between the peaks and between the troughs in the final stream length
are around $\Delta \Omega_{\rm b} \simeq \Omega_R/2 \simeq 13 \kms$,
which is reminiscent of equation (\ref{eq:resonant_condition}).
The similarity
between the 2D simulations and our Ophiuchus-like stream simulations
in terms of $\Delta \Omega_{\rm b}$
arises from the fact that
the pericentre locations with respect to the bar $\phi_{\rm rot}$
are important in shepherding the stream stars for both cases.
However, if we consider different 3D orbits,
different resonance conditions may be important.

\subsection{Comparison with observations} \label{section:observations}

Now that have created old streams which are short, we can compare them against observations of the Ophiuchus stream.
Our $N$-body simulations with $\Omega_{\rm b}/(\kmskpc)=-56.6$ and $69.1$
attain the final length of $2.0 \kpc$ and $1.0 \kpc$, respectively.
These lengths are comparable to the observed length of the Ophiuchus stream
($1.6 \pm 0.3 \kpc$; \citealt{Sesar2015}).
The case with $\Omega_{\rm b} = -56.6 \kmskpc$ is
especially interesting, since this value of $\Omega_{\rm b}$ is consistent with
a recent determination of $\Omega_{\rm b}$ \citep{Antoja2014}.

According to \cite{Sesar2015},
the width of the Ophiuchus stream is around $32 \pm 6 \pc$.
To evaluate the width of our $N$-body streams,
we calculated the distances of the stream stars from the orbital plane of the progenitor
and measured the dispersion of this distance at various positions along the stream.
We find that the width covering 68 \% of stream stars
is $9$-$31 \pc$ for $\Omega_{\rm b}=-56.6 \kmskpc$ model,
and $12$-$23 \pc$ for $\Omega_{\rm b}=-69.1 \kmskpc$ model.
Thus, these $N$-body models
are broadly consistent with the observed width of the Ophiuchus stream.

\begin{figure}
\begin{center}
\includegraphics[angle=0,width=0.95\columnwidth]{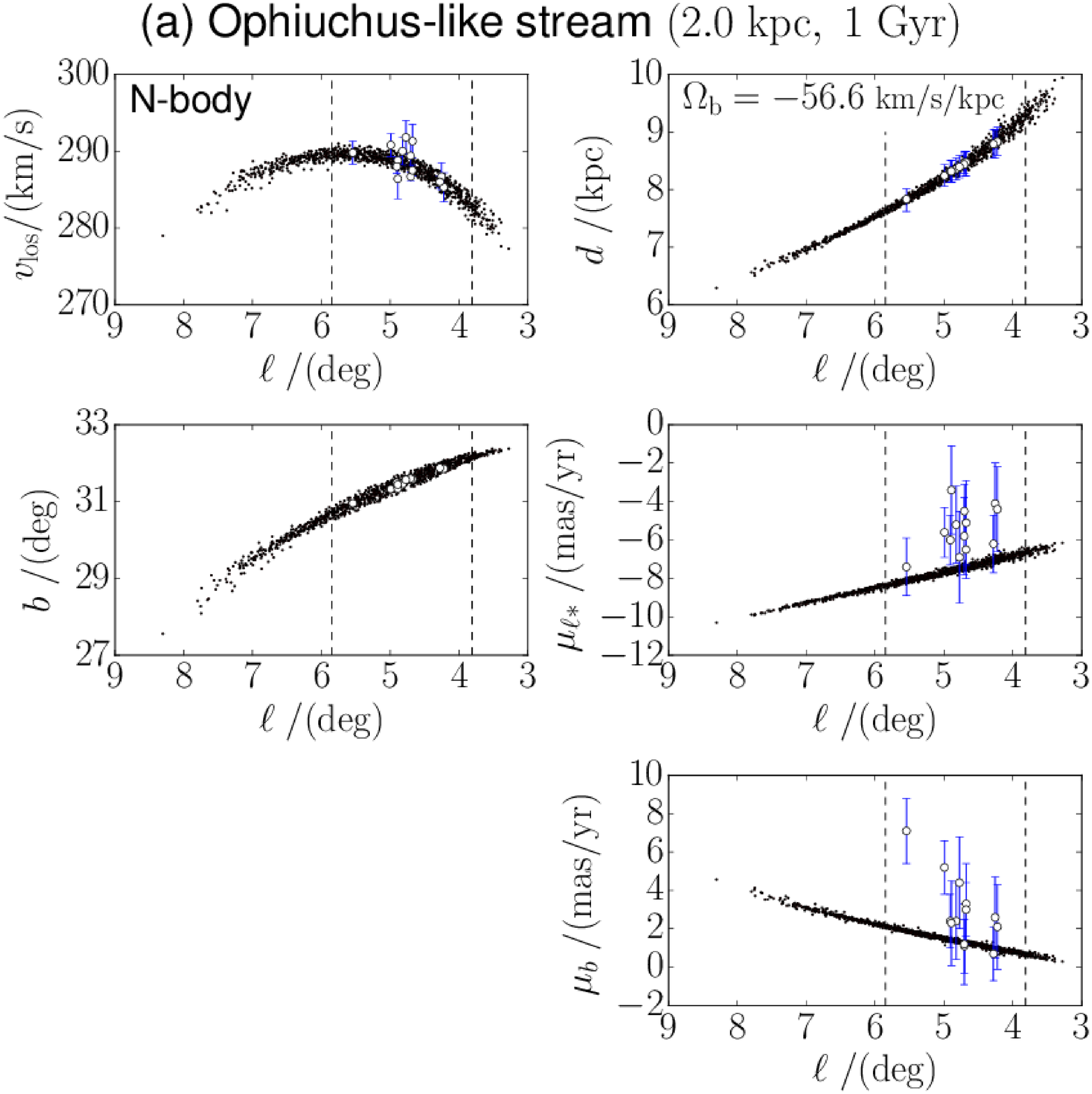}\\
\includegraphics[angle=0,width=0.95\columnwidth]{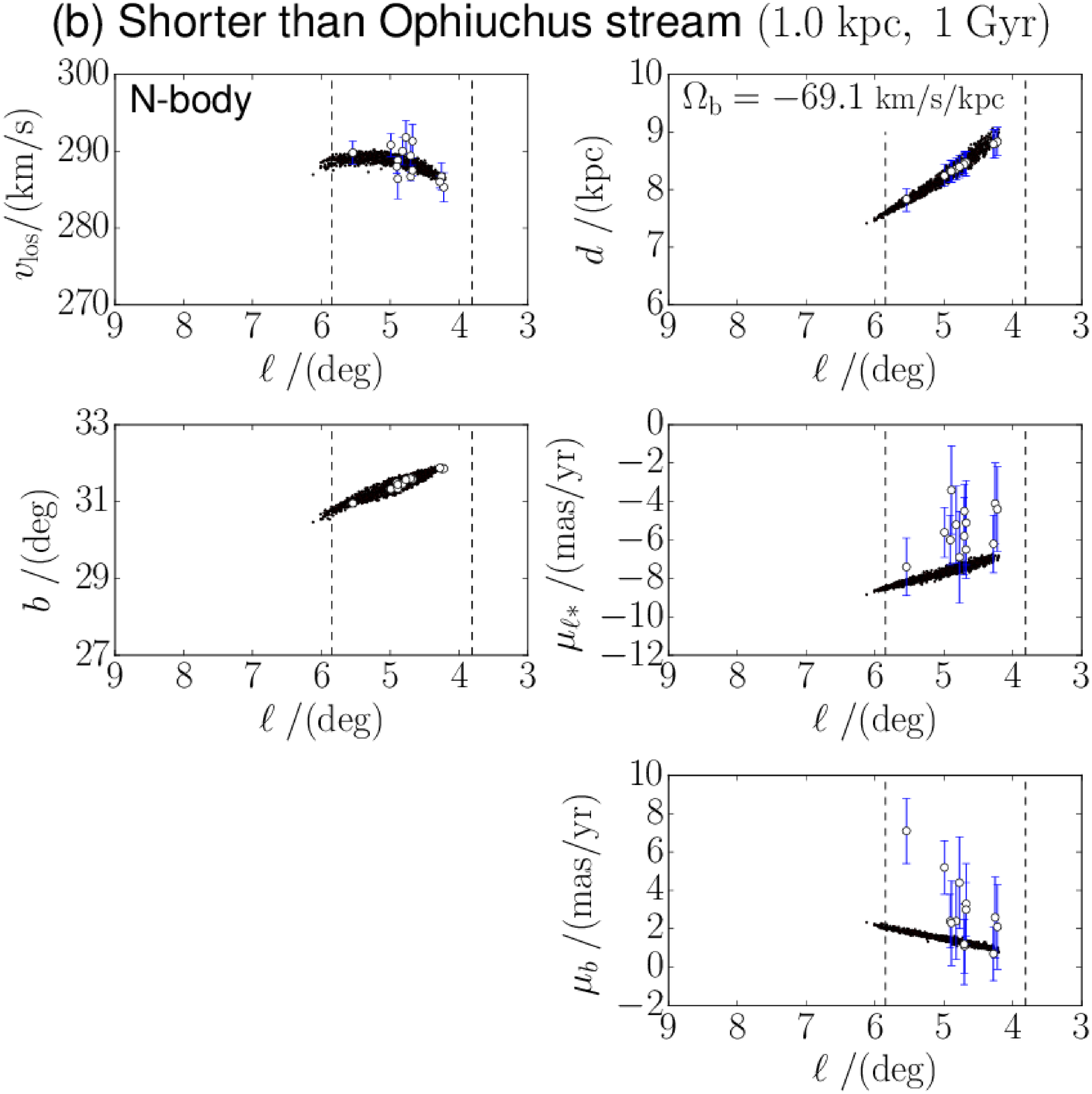}
\end{center}
\caption{
Comparison of our $N$-body simulations with observations of the Ophiuchus stream.
Here, $d$ denotes the heliocentric distance.
\textbf{(a)}
The top three rows show the results for $\Omega_{\rm b} = -56.6 \kmskpc$.
\textbf{(b)}
The bottom three rows show the results for $\Omega_{\rm b} = -69.1\kmskpc$.
In both cases, the bar rotates in the prograde direction ($\Omega_{\rm b}<0$).
The black dots represent the randomly-selected $10^3$ stars taken from our $N$-body results,
and the open circles (with or without error bar) represent the observed member stars of the Ophiuchus stream taken from \protect\cite{Sesar2015}.
The vertical dashed lines at $\ell=3.81^{\circ}$ and $5.85 ^{\circ}$
represent the edges of the stream described in \protect\cite{Sesar2015}.
}
\label{fig:Nbody}
\end{figure}

In Figure \ref{fig:Nbody} we compare our $N$-body models with observed data.
We see that
the observed spatial extent
and the line-of-sight velocities of the stream
are well reproduced in our models.
The proper motion data deviate from our models by design,
since we assume the proper motion of the progenitor (at $\ell=5 ^{\circ}$)
to be $(\mu_{\ell*}, \mu_b)=(-7.7, 1.5) \;{\rm mas \; yr^{-1}}$
so that the velocity vector aligns with the extent of the Ophiuchus stream as in \cite{Sesar2015}.
However, given the large errors on the proper motion,
our $N$-body models reasonably match the observations.

Our results suggest
that the Ophiuchus stream can be at least as old as $\sim1 \Gyr$
while keeping its length as short as $\sim 2 \kpc$,
depending on the bar's pattern speed.
This finding is in contrast to
the dynamical age of $\sim 300 \Myr$ claimed by \cite{Sesar2015}
who reached their conclusion using a static potential.

The values of $\Omega_{\rm b}$ that result in a short stream
depend on the assumed potential.
However, we have run additional Ophiuchus-like simulations in several trial potentials and
find that, for each trial potential,
there are several values of $\Omega_{\rm b}$
that result in short streams.
Thus,
it is not hard to construct a Ophiuchus-like stream model
that is as short as $\sim 2 \kpc$ with dynamical age of $\sim 1 \Gyr$,
if we assume that the stream has been evolved in a rotating barred potential.

\subsection{Uncertainties in proper motions and bar properties}
So far, we have focused on simulations
in which the bar properties and the final position and velocity of the progenitor are fixed.
Here we perform an additional set of Lagrange-point stripping simulations
to investigate how the observational uncertainties of both the proper motions of the Ophiuchus stream and the bar's pattern speed
affect the final length of the Ophiuchus-like stream.

For these additional simulations,
we fix the final location of the progenitor to be $(\ell, b)=(5.0, 31.37) ^{\circ}$ as before.
We assume that the distance modulus, line-of-sight velocity and the proper motions
follow Gaussian distributions
for which the mean and the dispersion are given by
$(\langle DM \rangle, \sigma_{DM}) = (14.57, 0.05)$,
$(\langle v_{\rm los} \rangle, \sigma_{v_{\rm los}}) = (289.1, 0.4) \kms$,
$(\langle \mu_{\ell*} \rangle, \sigma_{\mu_{\ell*}}) = (-7.7, 0.3) \;{\rm mas \; yr^{-1}}$
and
$(\langle \mu_b \rangle, \sigma_{\mu_b}) = (1.5, 0.3) \;{\rm mas \; yr^{-1}}$.
These dispersions are taken from Table 1 of \cite{Sesar2015}.
We randomly draw samples of the final conditions
based on these Gaussian distributions
and evolve the system for $\tau=1 \Gyr$.

\begin{figure}
\begin{center}
	\includegraphics[angle=0,width=0.95\columnwidth]{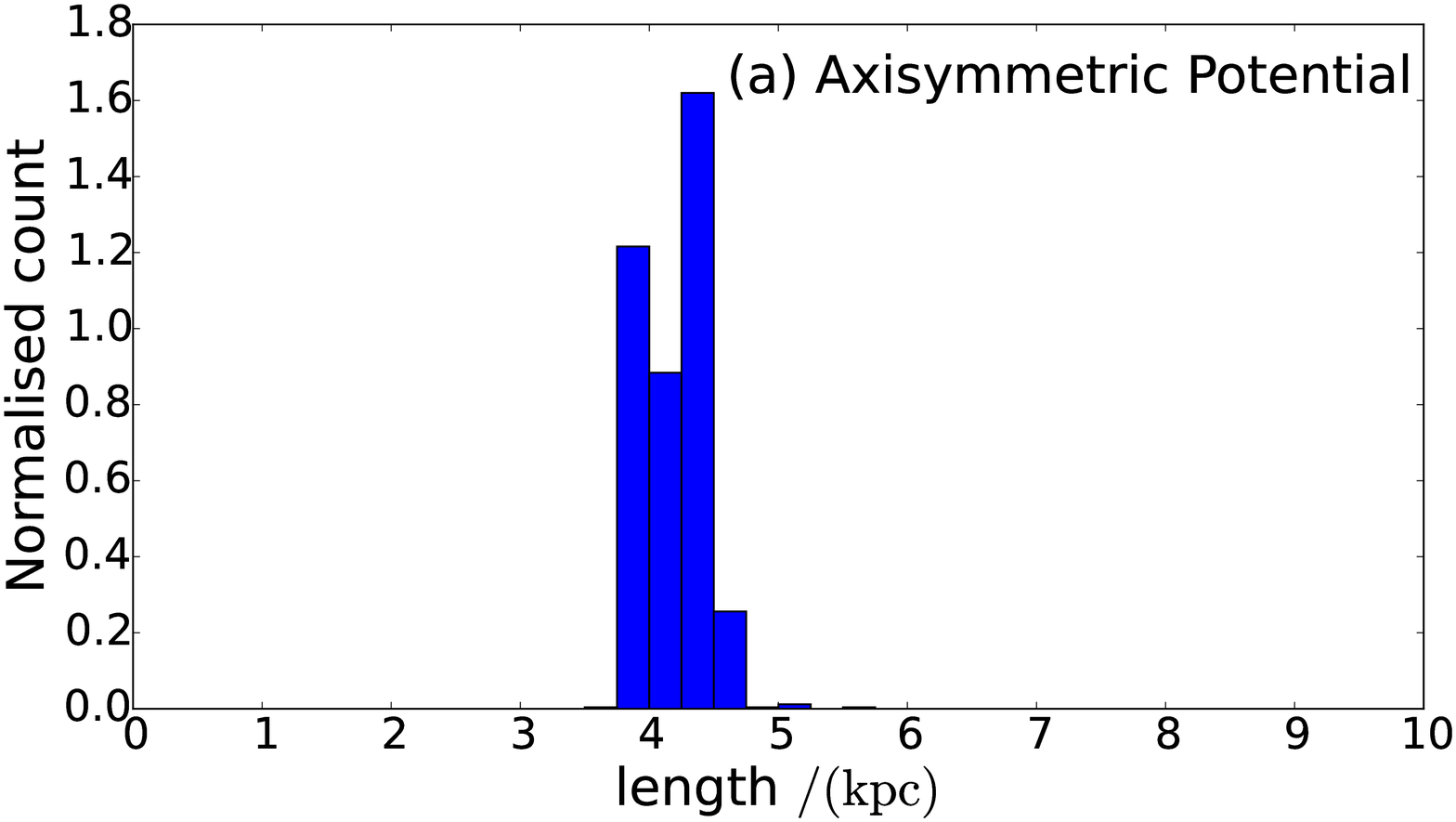}\\
	\includegraphics[angle=0,width=0.95\columnwidth]{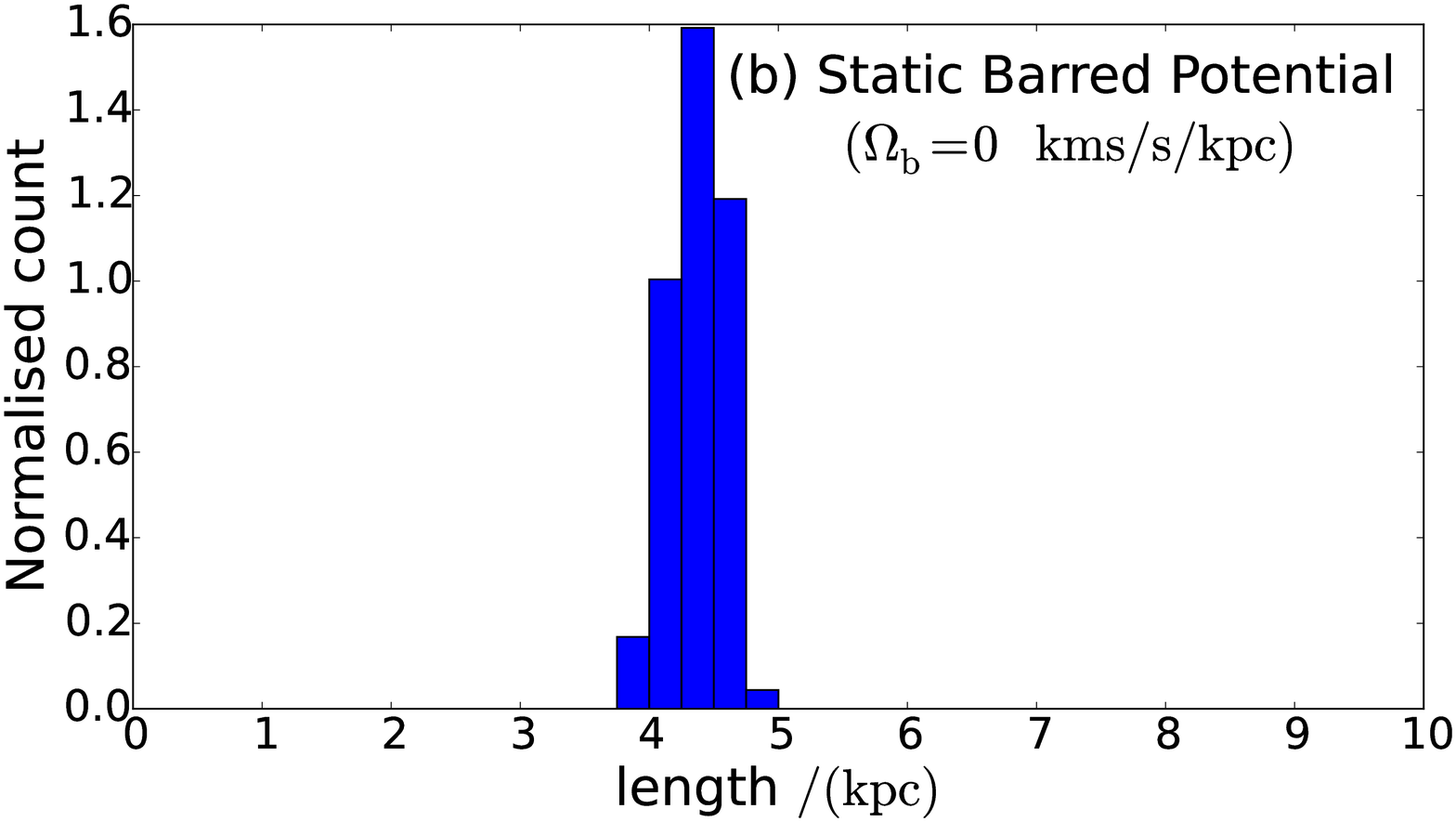}\\
	\includegraphics[angle=0,width=0.95\columnwidth]{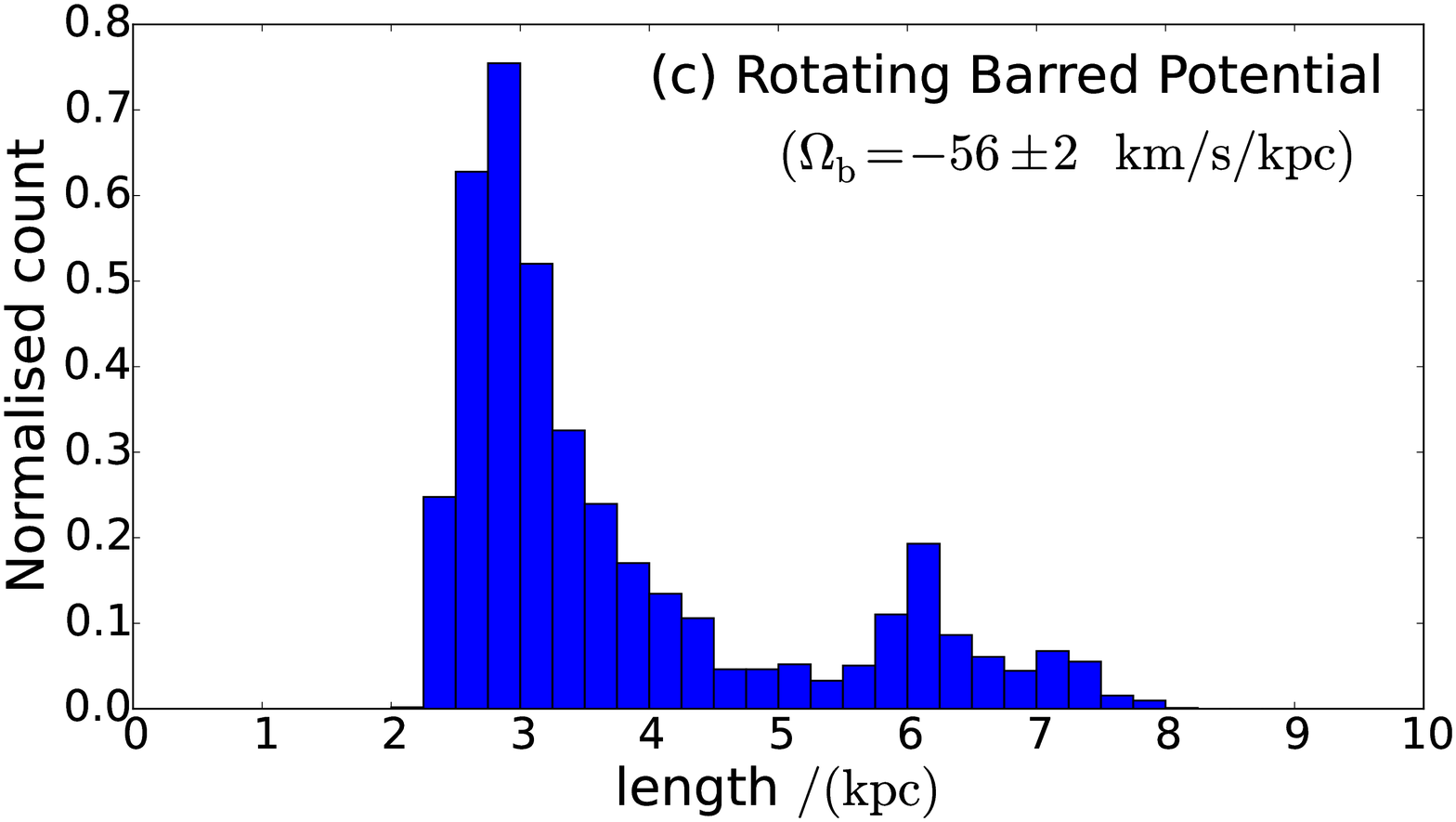}
\end{center}
\caption{
Probability distribution function of the final length of the Ophiuchus-like streams
when observational errors are taken into account.
\textbf{(a)} The case for the axisymmetric potential.
\textbf{(b)} The case for the static barred potential.
\textbf{(c)} The case for the rotating barred potential,
in which we assumed a Gaussian prior of $\Omega_{\rm b} = -56 \pm 2 \kmskpc$ \protect\citep{Antoja2014}.
Note that negative values of $\Omega_{\rm b}$ correspond to bars with prograde rotation.
}
\label{fig:histogram_length}
\end{figure}

First, we run simulations in the axisymmetric potential and the static barred potentials.
As seen in Figures \ref{fig:histogram_length}(a) and (b),
the final stream lengths for these cases
are narrowly distributed around $4 \kpc$
with a dispersion of $\sim 0.4 \kpc$.

Next, we run simulations in rotating barred potentials.
For each simulation,
we randomly assign the value of $\Omega_{\rm b}$
taken from a Gaussian prior of
$\Omega_{\rm b} = -56 \pm 2 \kmskpc$ \citep{Antoja2014}.
Figure \ref{fig:histogram_length}(c)
shows the distribution of the final length.
In this case,
we see a bimodal and wider distribution of the final length.
This bimodality means that
the stream length is squeezed by the bar for most of the runs,
while it is stretched for some of the runs.

\begin{figure}
\begin{center}
	\includegraphics[angle=0,width=0.95\columnwidth]{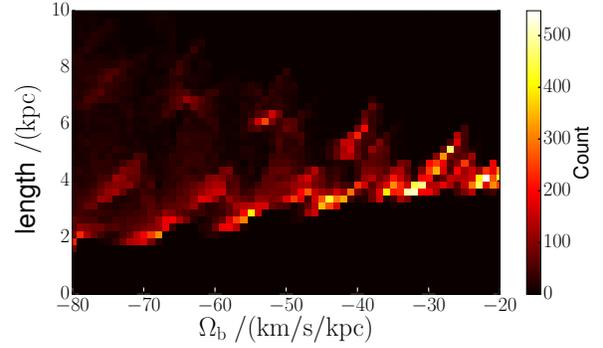}
\end{center}
\caption{
Probability distribution function of the final length of the Ophiuchus-like streams
as a function of $\Omega_{\rm b}$
when observational errors are taken into account.
Here we adopt a flat prior of $-80 \leq \Omega_{\rm b} / (\kmskpc) \leq -20$.
Note that negative values of $\Omega_{\rm b}$ correspond to bars with prograde rotation.
}
\label{fig:density_Omega_length}
\end{figure}

In order to see how the distribution of the final length depends on $\Omega_{\rm b}$,
we run $\simeq 6 \times 10^4$ simulations in the barred potential.
For each simulation,
we randomly sample $\Omega_{\rm b}$ from a flat prior of
$-80 \leq \Omega_{\rm b} / (\kmskpc) \leq -20$.
The distribution of the resultant simulations
in the $\Omega_{\rm b}$-length space
is shown in Figure \ref{fig:density_Omega_length}.
From this figure, we see that
the distribution of the final length is wider at larger $|\Omega_{\rm b}|$,
which is reminiscent of the larger oscillation in the final stream length at larger $|\Omega_{\rm b}|$
seen in Figure \ref{fig:3D_length_Omega}.
Also, we note that
a large fraction of the simulations end up with short streams ($\sim 2 \kpc$)
at $\Omega_{\rm b}/\kmskpc \simeq -60, -70$ and $-80$.

Figures \ref{fig:histogram_length} and \ref{fig:density_Omega_length}
provide important insights into the possible origin of the Ophiuchus stream.
If we assume that the Ophiuchus stream has been evolved for $1 \Gyr$ in a static potential (barred or not),
we expect the stream length is $\simeq4 \kpc$.
Thus, under this assumption,
one might reject a dynamical age of $\sim 1 \Gyr$ or older as in \cite{Sesar2015}.
However,
if we assume a rotating barred potential instead,
it is easier to explain the observed shortness of the Ophiuchus stream
even if its dynamical age is around $1 \Gyr$.
Moreover,
if we tune the potential model,
it might be possible to generate an Ophiuchus-like stream model
that is as short as the observed stream
but with a dynamical age significantly older than $1 \Gyr$.
Thus,
we argue that
the observed properties of the Ophiuchus stream
allow room for a dynamical age of $1 \Gyr$ or older.

\section{Discussion} \label{section:discussion}

\subsection{Lindblad Resonance}

At first sight, the mechanism described in this work seems similar to Lindblad resonance \citep{BT2008}. The Lindblad resonance occurs at orbital frequencies of
\eq{ n(\Omega_\phi - \Omega_{\rm b}) = \pm \Omega_R ,}
where $n$ is an integer. This should be contrasted with the resonant condition in this work from equation (\ref{eq:resonant_condition}) which we repeat here for clarity:
\eq{ (\Omega_\phi - \Omega_{\rm b}) = \pm \frac{N}{2} \Omega_R . }
The Lindblad resonance requires that resonant orbits are closed after one angular period in the co-rotating frame. Particles on nearby orbits are drawn to the resonant orbit and the density along the orbit is nearly uniform. Since stars are selected based on being near the resonant orbit, there can be a large scatter in age and metallicity of stars in the resonance. Finally, the higher order resonances (large $n$) are nearly circular.

In contrast, particles on orbits satisfying our resonant condition are required to approach the bar in the same orientation (modulo a rotation by 180$^\circ$) at each pericentre. These orbits can be much more eccentric than the orbits satisfying the Lindblad resonance, especially at large $N$. In addition, the density can vary significantly along the orbit since sections of the orbit which have pericentres near the centre of the bar will receive density enhancements, while those with pericentres near the major axis of the bar will have a depleted density.

We note that both the Lindblad resonance and our resonance condition are specific cases of the generalised condition in equation (\ref{eq:generalized_resonance}). The Lindblad resonance corresponds to $N=2$ with arbitrary $K$ and our resonance condition corresponds to $K=1$ with arbitrary $N$.

\subsection{Difference with chaotic fanning}

In recent work, \cite{Price-Whelan_et_al_2015} explored the effect of chaos on streams. They showed that chaos causes the stream to fan out since nearby orbits diverge exponentially in time. This fanning causes the stream density to drop off dramatically. \cite{Sesar2015b} proposed that if the Ophiuchus stream stars were on chaotic orbits, this mechanism could explain the relative shortness of the Ophiuchus stream. In order to distinguish whether chaotic fanning or the bar shepherding proposed in this work have contributed to the shortness of the Ophiuchus stream, we can compare their observational signatures. If the stream density has decreased due to fanning, we should expect to see an increase in the width of the stream towards the end of the Ophiuchus stream. We should also expect an increase in the velocity dispersion in the direction perpendicular to the stream plane. We note that this increase in dispersion may not necessarily be along the line of sight and may only be visible in the proper motion dispersion. In contrast to this, the bar shepherding mechanism does not appear to increase the dispersion or width near the end of the stream. In addition, better measurements of the bar's pattern speed and the Ophiuchus stream orbit will give a better constraint on exactly how close to resonance the stream is. Thus, we expect it should be possible to distinguish these mechanisms observationally.

\subsection{Implications for \emph{Gaia} and future surveys}
The arrival of the SDSS data produced a number of discoveries of substructure in the halo of the Milky Way \citep[e.g.][]{
Grillmair2006, Belokurov2007, GrillmairDionatos2006} and it is hoped that the next generation of surveys will produce many more such discoveries. Based on the SDSS discoveries it seems the most promising place to look for substructure is in the outer halo and it is traditionally held that the inner halo is composed of material that is already well phase-mixed \citep{Helmi2008}. However, the mechanism described in this work can either increase or decrease the phase-space density of tidal debris which comes near the bar. As a result, tidal debris which would otherwise have phase-mixed in the Milky Way potential can remain visible for long periods of time. Since the effect depends sensitively on the phase at which the debris approaches the bar, we expect enhanced debris will belong to a single progenitor as opposed to the variety of stars seen in Lindblad resonances. Therefore, although stream-hunting in the outer halo is expected to be more profitable, stream-hunting within the central regions of the Galaxy is also anticipated to bear fruit.
With this regard, it is worthwhile pointing out that an infrared satellite \emph{JASMINE} (Japan Astrometry Satellite Mission for INfrared Exploration; \citealt{Gouda2012}) plans to perform an astrometric survey of the bulge in 2020s.

Our simulations have focussed on potentials with bars that rotate with a constant pattern speed. $N$-body simulations of discs embedded in halos have demonstrated that the bar can be torqued by the halo such that it gradually slows over time \citep{DebattistaSellwood2000, MartinezValpuesta2006}. For our model of the Ophiuchus stream to be valid we require the pattern speed of the bar to change by only a few $\kmskpc$ over a $\Gyr$.
This required rate of change in $|\Omega_{\rm b}|$ is comparable
to the rate seen in numerical simulations (see Figure 2 of \citealt{MartinezValpuesta2006}).
Significant evolution of the pattern speed would cause the progenitor orbit to shift from near a shortening resonance to near a lengthening resonance and would begin to wash out the signal we have observed. However, this situation also gives us hope that segments of longer streams may become resonantly trapped and be observed as overdensities in much longer less dense streams. Optimistically, we might hope that observations of many short streams near the centre of the Galaxy may provide constraints on the time evolution of the Galactic bar.

\section{Conclusion} \label{section:conclusion}
We have identified a novel mechanism which can modify the growth rate of streams which pass near the bar. Since different parts of a stream arrive at pericentre at different times, they receive different torques and hence different changes in energy. This in turn modifies the orbital frequencies of each particle, which changes the growth rate of the stream. We have demonstrated that for streams confined to the equatorial plane of an axisymmetric potential plus a rotating bar, the length of the stream is a strong function of the bar pattern speed. When the pattern speed is such that the progenitor approaches the minor axis of the bar at each pericentric passage, the resulting stream is much longer than if the stream were evolved in a completely axisymmetric potential. However, when the pericentric passages of the progenitor are along a line perpendicular to the bar's major axis the opposite effect is seen and the resulting stream is significantly shorter.

This mechanism allows the existence of short and old stellar streams.
We have run a series of Lagrange-point stripping method simulations as well as $N$-body simulations
and investigated how a rotating bar affects the length of the Ophiuchus stream.
We find that some of our Ophiuchus-like $N$-body streams evolved in a barred potential for $1 \Gyr$
can be as short as 1-2 $\kpc$ and can explain the observed properties,
if the pattern speed of the bar is properly chosen.
This finding suggests that the dynamical age of the Ophiuchus stream can be at least as long as $\sim 1 \Gyr$.
This age is in contrast to the dynamical age of $\sim 300 \Myr$ claimed by \cite{Sesar2015}, who assumed an axisymmetric potential.
We note that
it might be possible to generate Ophiuchus-like stream models with dynamical ages much older than $1 \Gyr$
by tuning the potential model.

Our mechanism opens a new possibility of unveiling the formation history of the inner halo.
Previously, it has been thought that the inner halo of the Milky Way is highly phase-mixed due to the short dynamical time \citep{Helmi2008}.
This conventional view suggests that the spatial and velocity distribution of inner halo stars are already smoothed out
and have lost coherent structures that retain the initial conditions of the accreted stellar systems.
However, our mechanism can enhance the density of some stellar streams in the inner halo.
With future astrometric surveys near the bulge region (\emph{Gaia}) or within the bulge (\emph{JASMINE}),
we might be able to detect more substructures similar to the Ophiuchus stream in the inner halo.
Such inner halo substructures will be beneficial in understanding
the accretion history of the Milky Way as well as the dynamical history of the Galactic bar.

\section*{Acknowledgments}
The authors thank Wyn N. Evans for helpful comments on the manuscript. 
KH is supported by Japan Society for the Promotion of Science (JSPS) through Postdoctoral Fellowship for Research Abroad. The research leading to these
results has received funding from the European Research Council under
the European Union's Seventh Framework Programme (FP/2007-2013)/ERC
Grant Agreement no. 308024. JLS acknowledges the support of the Science and Technology Facilities Council. All authors thank the valuable discussions with the Cambridge streams group throughout the course of this work.

\appendix

\section{Shepherding the distribution of stream stars}

\begin{figure*}
\begin{center}
	\includegraphics[angle=0,width=0.7\textwidth]{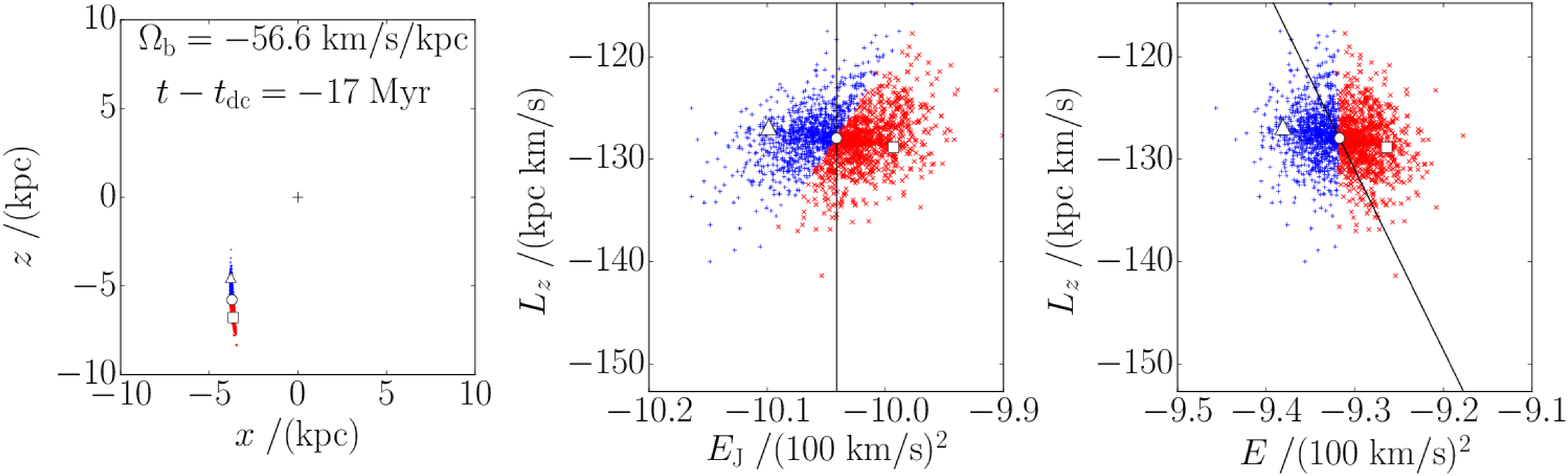}\\
	\includegraphics[angle=0,width=0.7\textwidth]{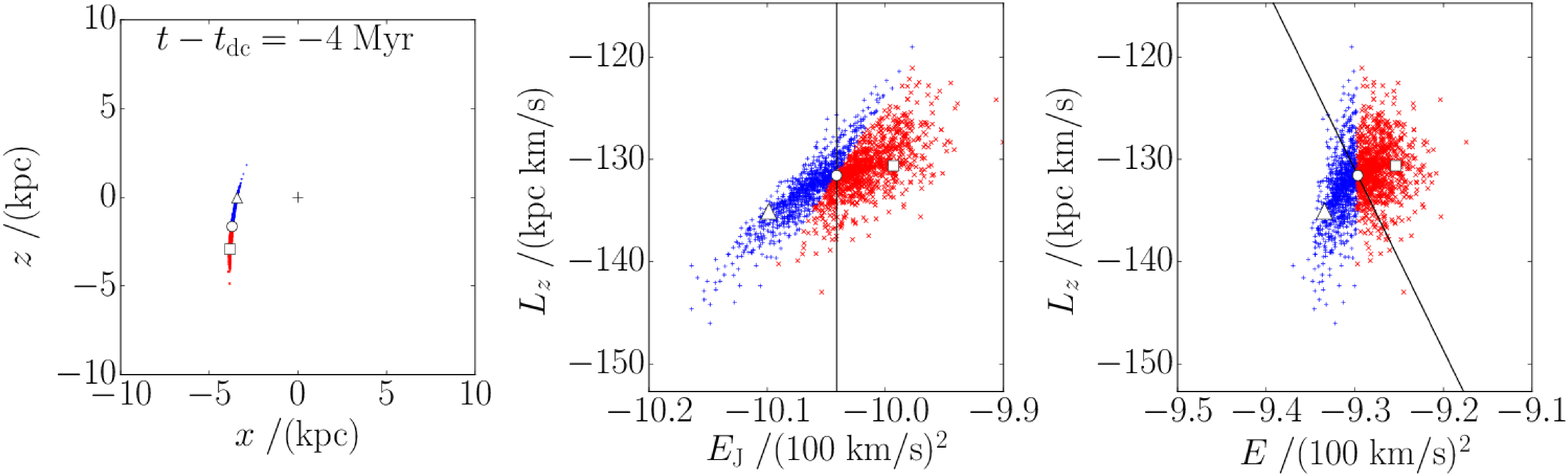}\\
	\includegraphics[angle=0,width=0.7\textwidth]{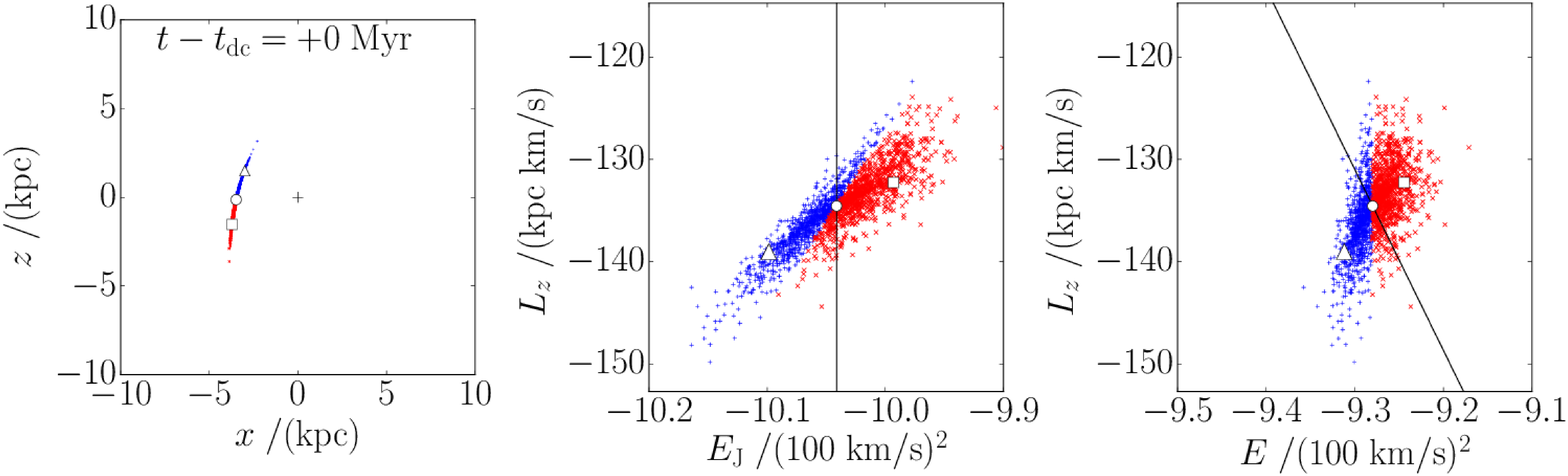}\\
	\includegraphics[angle=0,width=0.7\textwidth]{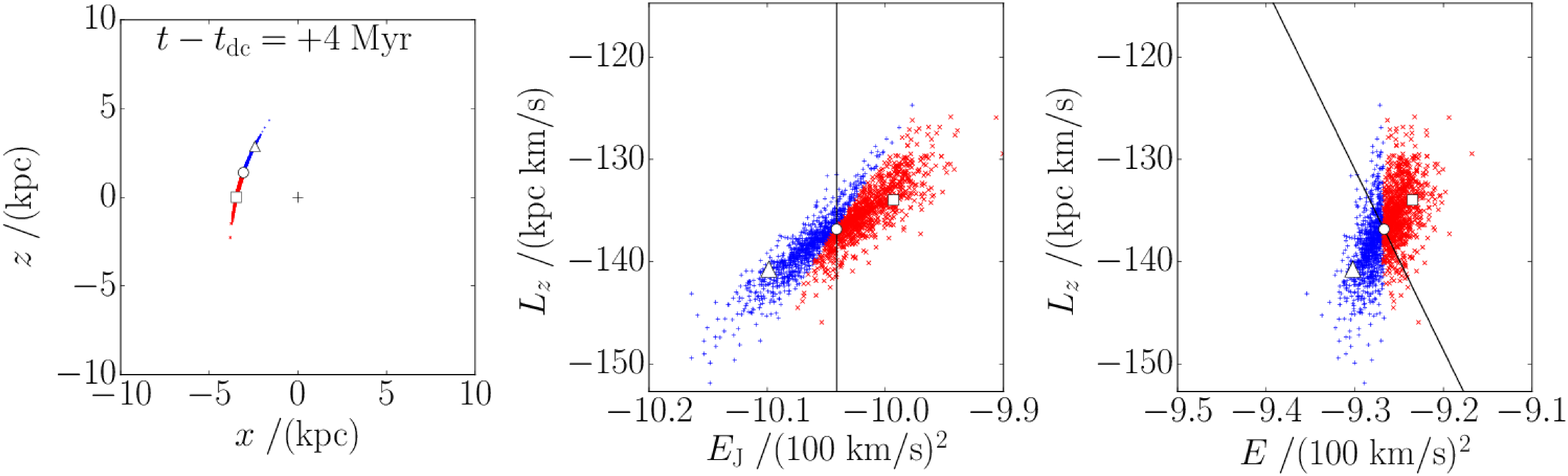}\\
	\includegraphics[angle=0,width=0.7\textwidth]{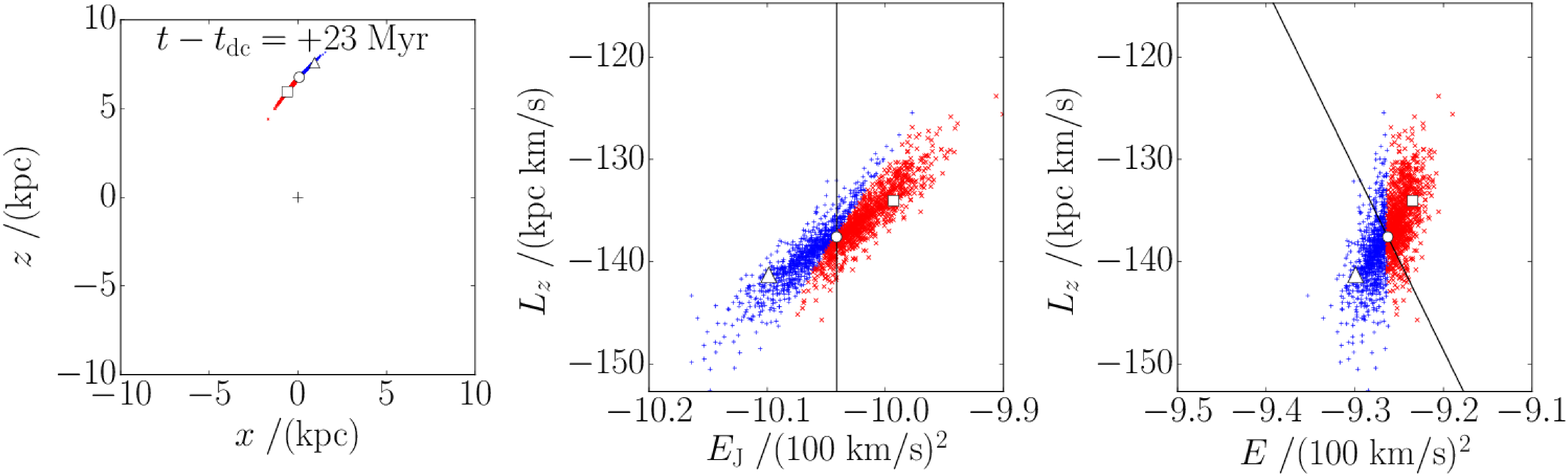}
\end{center}
\caption{
The time-evolutions of the distribution of $E$, $L_z$ and $E_{\rm J}$ of stream stars in the same simulation as in Figure \ref{fig:3D_special_cases_track}.
The different rows represent the distribution of stars
at different time epochs
with respect to the second disc crossing time $(t=t_{\rm dc}\equiv -0.743 \Gyr)$ of the progenitor
(time flows from top to bottom row).
During this disc crossing, the progenitor is moving from $z<0$ towards $z>0$.
The left-hand column shows the distribution of stars in the $(x,z)$-plane.
The middle and right columns shows the distributions of stars in the $(E_{\rm J}, L_z)$- and $(E, L_z)$-space.
The blue `$+$' and red `$\times$' represent the leading and trailing stars, respectively.
The open circle represents the progenitor system.
The open triangle and open square
represent the the leading and trailing stars shown in Figure \ref{fig:3D_special_cases_track}.
The vertical line in the middle panels and the inclined line in the right panels
show the lines of constant $E_{\rm J}$.
Due to the conservation of $E_{\rm J}$,
all the particles move parallel to these lines.
}
\label{fig:3D_special_cases_distribution}
\end{figure*}

In this appendix we investigate how the distribution of $(E,L_z)$ of stream stars changes as
the stream crosses the disc plane. In Figure \ref{fig:3D_special_cases_distribution} we show
the distribution of stream stars in the $\Omega_{\rm b}=-56.6 \kmskpc$ simulation
during the same disc-crossing as in Figure \ref{fig:3D_special_cases}.
The left, middle and right columns
show respectively
the snapshot
in the inertial $(x, z)$-plane,
in the $(E_{\rm J}, L_z)$-plane, and
in the $(E, L_z)$-plane.
The time evolves from the top row to the bottom,
and the stream is moving towards $(+z)$-direction.
The leading star and the trailing star
that we have probed in Figure \ref{fig:3D_special_cases}
are marked by the open triangle and square, respectively;
while the progenitor is marked by the open circle.
Those stars in the leading and trailing tails are shown by
blue $+$ and red $\times$, respectively.
We note that the each star conserves $E_{\rm J}$,
so that each star moves only
vertically on the middle panels
and in parallel to the inclined line on the right panels. Finally, in each row we give the time relative to the disc crossing time, $t_{\rm dc}$.

At $t-t_{\rm dc}=-17 \Myr$,
the stream is well below the Galactic plane
and we see that the leading-tail stars have larger $L_z$ than the trailing-tail stars on average.

At $t-t_{\rm dc}=-4 \Myr$,
the leading tail reaches the disc plane.
Here, the leading-tail stars decrease $L_z$ by a large amount.
In contrast, the distribution of $L_z$ does not change a lot in the trailing tail
since the trailing tail is yet too far from the disc plane.
As a result, the distributions in the $(E_{\rm J}, L_z)$- and $(E, L_z)$-plane
become notably asymmetric.
This asymmetry can be well traced
by looking at the relative position of the triangle and the square.

At $t=t_{\rm dc}$, the progenitor reaches the disc plane,
and subsequently the trailing tail crosses the disc plane at $t-t_{\rm dc}=14 \Myr$.
By the time the trailing tail experiences the disc crossing,
the bar's influence on the stream is no longer strong,
so that the change in $L_z$ is mild for the trailing-tail stars.
Thus, the distribution of the trailing-tail stars
in the $(E_{\rm J}, L_z)$- and $(E, L_z)$-plane
is nearly frozen at $0 \Myr < t-t_{\rm dc} < 23 \Myr$.

As a result,
the correlation between $E$ and $L_z$ of the stream stars
is changed during this disc crossing.
This is the fundamental reason why we
see a sharp increase in $\sigma(L_z)$ and sharp decrease in $\sigma(E)$
at the second disc crossing of the $\Omega_{\rm b}=-56.6 \kmskpc$ simulation.

\label{lastpage}

\end{document}